\documentclass[]{emulateapj}

\def\mycite#1{\citeauthor{#1} \citeyear{#1}}
\def\deg{\hbox{$^\circ$}}
\def\eff {{\cal E}}
\def\rate{\eta}
\def\exposure{{\it E}}

\def\myeps@scaling{.55}
\def\myplotone#1{\centering \leavevmode
    \epsfxsize=\myeps@scaling\columnwidth \epsfbox{#1}}

\newcommand{\arcs}{{$^{\prime\prime}\,$}}

\begin{document}

\submitted{Accepted, \apj~10 August 2004}

\title{The Deep Lens Survey Transient Search I : Short Timescale and Astrometric Variability}
\author{
  A.C.~Becker\altaffilmark{1,2,3},
  D.M.~Wittman\altaffilmark{1,4},
  P.C.~Boeshaar\altaffilmark{4,5},
  A.~Clocchiatti\altaffilmark{6},
  I.P.~Dell'Antonio\altaffilmark{7},
  D.A.~Frail\altaffilmark{8},
  J.~Halpern\altaffilmark{9},
  V.E.~Margoniner\altaffilmark{1,4},
  D.~Norman\altaffilmark{10},
  J.A.~Tyson\altaffilmark{1,4},
  R.A.~Schommer\altaffilmark{$\dagger$}
}

\altaffiltext{1}{Bell Laboratories, Lucent Technologies, 600 Mountain Avenue, Murray Hill, NJ 07974\\
  Email: {\tt acbecker,dwittman,tyson,vm21@lucent.com}}
\altaffiltext{2}{NIS-2, Space and Remote Sciences Group, Los Alamos National Laboratory, Los Alamos, NM 87545}
\altaffiltext{3}{Astronomy Department, University of Washington, Seattle, WA 98195}
\altaffiltext{4}{Physics Department, University of California, Davis, CA 95616 \\
  Email: {\tt boeshaar,margoniner,tyson,wittman@physics.ucdavis.edu}}
\altaffiltext{5}{Drew University, Madison, NJ, 07940 \\
  Email: {\tt pboeshaa@drew.edu}}
\altaffiltext{6}{Pontificia Universidad Cat\'{o}lica de Chile,
  Departamento de Astronom\'{i}a y Astrof\'{i}sica, Casilla 306,
  Santiago 22, Chile \\
  Email: {\tt aclocchi@astro.puc.cl}}
\altaffiltext{7}{Physics Department, Brown University, Providence, R.I. 02912 \\
  Email: {\tt ian@het.brown.edu}}
\altaffiltext{8}{National Radio Astronomy Observatory, P.O. Box 'O', Socorro, NM 87801 \\
  Email: {\tt dfrail@nrao.edu}}
\altaffiltext{9}{Department of Astronomy, Columbia University, New York, NY, 10025-6601 \\
  Email: {\tt jules@astro.columbia.edu}}
\altaffiltext{10}{Cerro Tololo Inter--American Observatory, National Optical Astronomy Observatory, Casilla 603, La Serena, Chile \\
  Email: {\tt dnorman@ctio.noao.edu}}
\altaffiltext{$\dagger$}{Deceased 2001 December 12}

\begin{abstract} 
We report on the methodology and first results from the Deep Lens
Survey (DLS) transient search.  We utilize image subtraction on survey
data to yield all sources of optical variability down to 24$^{th}$
magnitude.  Images are analyzed immediately after acquisition, at the
telescope and in near--real time, to allow for followup in the case of
time--critical events.  All classes of transients are posted to the
web upon detection.  Our observing strategy allows sensitivity to
variability over several decades in timescale.  The DLS is the first
survey to classify and report all types of photometric and astrometric
variability detected, including solar system objects, variable stars,
supernovae, and short timescale phenomena.
Three unusual optical transient events were detected, flaring on
thousand--second timescales.  All three events were seen in the $B$
passband, suggesting blue color indices for the phenomena.  One event
(OT 20020115) is determined to be from a flaring Galactic dwarf star
of spectral type dM4.  From the remaining two events, we find an
overall rate of $\rate = 1.4~ {\rm events}~ {\rm deg}^{-2}~{\rm
day}^{-1}$ on thousand--second timescales, with a $95\%$ confidence
limit of $\rate < 4.3$.  One of these events (OT 20010326) originated
from a compact precursor in the field of galaxy cluster Abell 1836,
and its nature is uncertain.  For the second (OT 20030305) we find
strong evidence for an extended extragalactic host.  A dearth of such
events in the $R$ passband yields an upper $95\%$ confidence limit on
short timescale astronomical variability between $19.5 < \mathcal{M}_R
< 23.4$ of $\rate_R < 5.2~ {\rm events}~ {\rm deg}^{-2}~{\rm
day}^{-1}$.  We report also on our ensemble of astrometrically
variable objects, as well as an example of photometric variability
with an undetected precursor.

\end{abstract}
\keywords{gamma rays: bursts --- minor planets, asteroids --- stars: variables: other --- supernovae: general --- surveys}

\section{Optical Astronomical Variability}
\label{sec-intro}

Characterization of the variable optical sky is one of the new
observational frontiers in astrophysics, with vast regions of
parameter space remaining unexplored.  At the faint flux levels
reached by this 
optical transient search, previous surveys were only able to probe
down to timescales of hours.  An increase in observational sensitivity
at short timescale and low peak flux holds the promise of detection
and characterization of rare, violent events, as well as new
astrophysics.

The detection of transient optical emission provides a window into a
range of known astrophysical events, from stellar variability and
explosions to the mergers of compact stellar remnants.  Known types of
catastrophic stellar explosions, such as supernovae and gamma--ray
bursts (GRBs), produce prompt optical transients decaying with
timescales of hours to months.  Some classes of GRBs result from the
explosion of massive stars -- these hypernovae are known to produce
bright optical flashes decaying with hour--long timescales due to
emission from a reverse shock plowing into ejecta from the explosion.
In addition, GRBs produce optical afterglows, decaying on day to week
timescales, resulting from jet--like relativistic shocks expanding
into a circumstellar medium.  Even more interesting are explosive
events yet to be discovered, such as mergers among neutron stars and
black holes.  These may have little or no high-energy emission, and
hence may be discoverable only at longer wavelengths \citep{li-pac}.
Finally, there is the opportunity to find rare examples of
variability, as well as the potential for discovering new,
unanticipated phenomena.

One of the primary science goals of this transient search is the rate
and distribution of short timescale astronomical variability.
However, we catalogue and report all classes of photometric or
astrometric variability with equal consideration.  A weakness in many
surveys with targeted science is that serendipitous information is
discarded as background -- a counterexample is the wealth of stellar
variability information being gleaned from microlensing survey data.
The DLS survey geometry and observation cadence are not optimal for
maximizing the overall number of detected transients
(e.g. \mycite{nemiroff-03}), being driven instead by weak lensing
science requirements.  However, given the current lack of constraints
at short timescales and to significant depth, the potential remains
for the discovery of new types of astronomical variability.

The search for variability without population bias is one of the
primary goals of the Large-aperture Synoptic Survey Telescope (LSST,
\mycite{tyson-lsst02}).  Efforts such as the DLS transient search are
a useful and necessary testing ground for the software design and
implementation needed to reduce, in real--time, LSST's expected data
flow of 20 TB nightly.

\section{Variability Surveys}
\label{sec-varhist}

The field of optical variability surveys has been revolutionized by
the adaption of CCD devices to astronomy, and by Moore's progression
of computing power.  The former has yielded wide field imaging systems
on large aperture telescopes, and the latter ensures their data can be
reduced in quick order, if not real--time.  Such advances have lead to
modern variability surveys characterized by successively more extreme
combinations of depth, cadence, and sky coverage.

The advent of microlensing surveys advanced by orders of magnitude the
known number of variable objects, driving the forefront of wide--field
(0.1--1 deg$^2$) imaging on meter--class telescopes (e.g. EROS:
\mycite{eros-smc03}; MACHO: \mycite{macho-lmc2000}; MEGA:
\mycite{mega-03}; MOA: \mycite{moa-bulge03}; OGLE I,II,III:
\mycite{ogle-events01}; POINT-AGAPE: \mycite{pointagape-03};
SuperMACHO: \mycite{supermacho-03}).  The cadences are typically of
order 1 image per day per field, although various strategies have been
employed for sensitivity to short events or fine structure in longer
events.  Such surveys are traditionally directed at crowded systems of
stars serving to backlight a foreground microlensing population.  This
stellar background, however, also serves as a foreground that
obfuscates sources at cosmological distances.

Surveys for supernovae at cosmological distances avoid such crowded
foregrounds, employing similar resources and attaining similar sky
coverage, but reaching deeper than the microlensing searches
(e.g. High-z: \mycite{tonry-highz}; SCP: \mycite{scp-03}).  The
original surveys' cadence of field reacquisition was typically of
order a month, short enough to avoid confusion by active galactic
nuclei but long enough to ensure a handful of new supernova per
observing run.  These surveys were designed only for event discovery,
with subsequent lightcurve coverage provided by narrower field
followup resources.  Given the successes of the original supernovae
surveys, subsequent efforts have been allocated resources for an
advanced observing cadence allowing both detection and self--followup,
such as the IfA Deep \citep{barris-ifa} and ESSENCE \citep{essence-02}
supernova surveys.  The Nearby Supernova Factory is expecting $\sim
100$ type Ia supernova per year from their moderate aperture
(1.2--meter) telescope and large (9 deg$^2$) field of view
\citep{snfac}.  However, dedicated systems with a small field of view
(0.01 deg$^2$), similar depth, and a cadence of several days can lead
to a wealth of supernova detections (e.g. LOSS: \mycite{loss}).

Asteroid and Near Earth Object (NEO) searches cover of order thousands
of degrees per night, primarily in the ecliptic, and near opposition
(e.g. LONEOS: \mycite{loneos}; LINEAR: \mycite{linear}; NEAT:
\mycite{neat-asteroid}).  Their photometric depths are shallower than
supernova surveys and the cadence more rapid, given the requirement of
recovering objects in motion.  The images are typically of poorer
seeing and image sampling, given less stringent photometric
requirements.

Truly wide--field systems (10 to 100 deg$^2$) incorporating small
apertures and limiting survey depths of 12--17$^{th}$ magnitudes are
able to cover the entirety of the sky in a given night, or a given
large region of sky multiple instances per night (e.g. GROSCE:
\mycite{grocse}; ROTSE-III: \mycite{rotse-3}; RAPTOR: \mycite{raptor};
ASAS-3: \mycite{asas-3}).  Such systems are optimized for the
detection and followup of fast optical transients, such as optical
counterparts to GRBs.  These systems compromise Nyquist sampling of
the point spread function (PSF) and photometric depth for breadth and
real--time agility.  RAPTOR's advanced implementation includes a
narrow--field fovea to optimally study localized transients.  The
Northern Sky Variability Survey (NSVS: \mycite{nsvs}) makes use of
ROTSE-I survey data, which sampled the entire local sky up to twice
nightly and provides variability information on $\sim 14$ million
objects between 8--15.5$^{th}$ magnitudes.

Additional variability surveys include the QUEST RR Lyrae survey
\citep{vivas-quest}, which is to cover 700 deg$^2$ to $V = 21^{st}$
magnitude in the Galactic equator, with sensitivity to several
timescales of variability.  The Sloan Digital Sky Survey has a
sequence of multi--color temporal information for over 700 deg$^2$ at
multiple timescales \citep{ivesic-sdsstemporal}.  The Palomar-Quest
synoptic sky survey drift--scans 500 deg$^2$ per night, with scans
separated by time baselines of days to months \citep{palomar-quest}.
The Deep Ecliptic Survey monitors 13 deg$^2$ per night near the
ecliptic in a search for Kuiper Belt Objects (KBOs), using 4--meter
telescopes and 2 exposures separated by 2--3 hours and a third within
a day \citep{des}.  Finally, the Faint Sky Variability Survey (FSVS:
\mycite{fsvs-03}) is a study of overall optical and astrometric
variability at faint magnitudes on a 2.5--meter telescope.  Point
source limiting magnitudes reach to 25 over a field of $\sim 0.1$
deg$^2$, and the cadence of field reacquisitions is staggered for
sensitivity at several timescales, and thus to multiple phenomena.  Of
all on--going variability surveys, the FSVS is most similar to the DLS
transient search.

\section{The Deep Lens Survey}
\label{sec-survey}

The Deep Lens Survey (DLS) is a 5--year NOAO survey operating on the
4--meter Blanco and Mayall telescopes + MOSAIC imagers at the Cerro
Tololo and Kitt Peak observatories, respectively \citep{dls-spie02}.
The survey is undertaking very deep multicolor imaging of five $2 \deg
\times 2 \deg$ fields chosen at high Galactic latitude, and composed
of 9 subfields apiece.  Exposure times and filters are chosen to reach
limiting magnitudes of $B,V,R,z'$ to $29/29/29/28$ mag per square
arcsecond surface brightness, with images typically acquired near new
moon.  At these limiting magnitudes, we expect to measure accurately
the shapes and color--redshifts of $\sim 100,000$ galaxies per square
degree for weak lensing science.  An overview of the survey and public
data releases are available at \url{http://dls.bell-labs.com/}.  The
survey is expected to conclude observations in March, 2005.

Full exposure depth requires 20 exposures per passband in each of our
45 subfields.  The images are obtained, with some exceptions, in sets
of 5--pointing dither sequences, with offsets of $100\arcsec$ and
$200\arcsec$ around the initial central pointing to fill gaps between
MOSAIC chips and enable super--sky--flat construction.  Typical
exposure times are 600 seconds in $B$, $V$, and $z'$, and 900 seconds
in $R$, yielding limiting magnitudes per exposure for point sources of
approximately $24$.  The weak lensing science goals of the survey
require co--adding these images.  In the co--addition process,
however, variable sources are clipped out, or averaged away.  Here we
pursue an orthogonal approach, which instead analyzes {\it
differences} in these data, at the pixel level, to extract this
variability.  This paper summarizes our transient search to date.

\section{DLS Transient Search}
\label{sec-trans}

Starting in December 1999, we began analysis of our acquired images in
near--real time, which we define as within the duration of a dither
sequence of observations.  True real--time reductions ideally occur
within the timescale it takes to acquire the subsequent observation,
or, given a target population of events, shorter than the timescale of
the variability being searched for.  
Recognizing variability in near--real time with current CCD mosaics
requires data analysis of several GB nightly of images on--site.  To
this end, we maintain computers in the control rooms of both the Kitt
Peak and Cerro--Tololo 4--meter telescopes.  At Kitt Peak, we have a
quad--processor 700 MHz machine, and at Cerro--Tololo a
quad--processor 550 MHz machine.  Both systems run under the Linux
operating environment, and execute the custom data reduction pipeline
described in Section~\ref{sec-pipeline}.

\begin{figure*}[t]
  \plotone{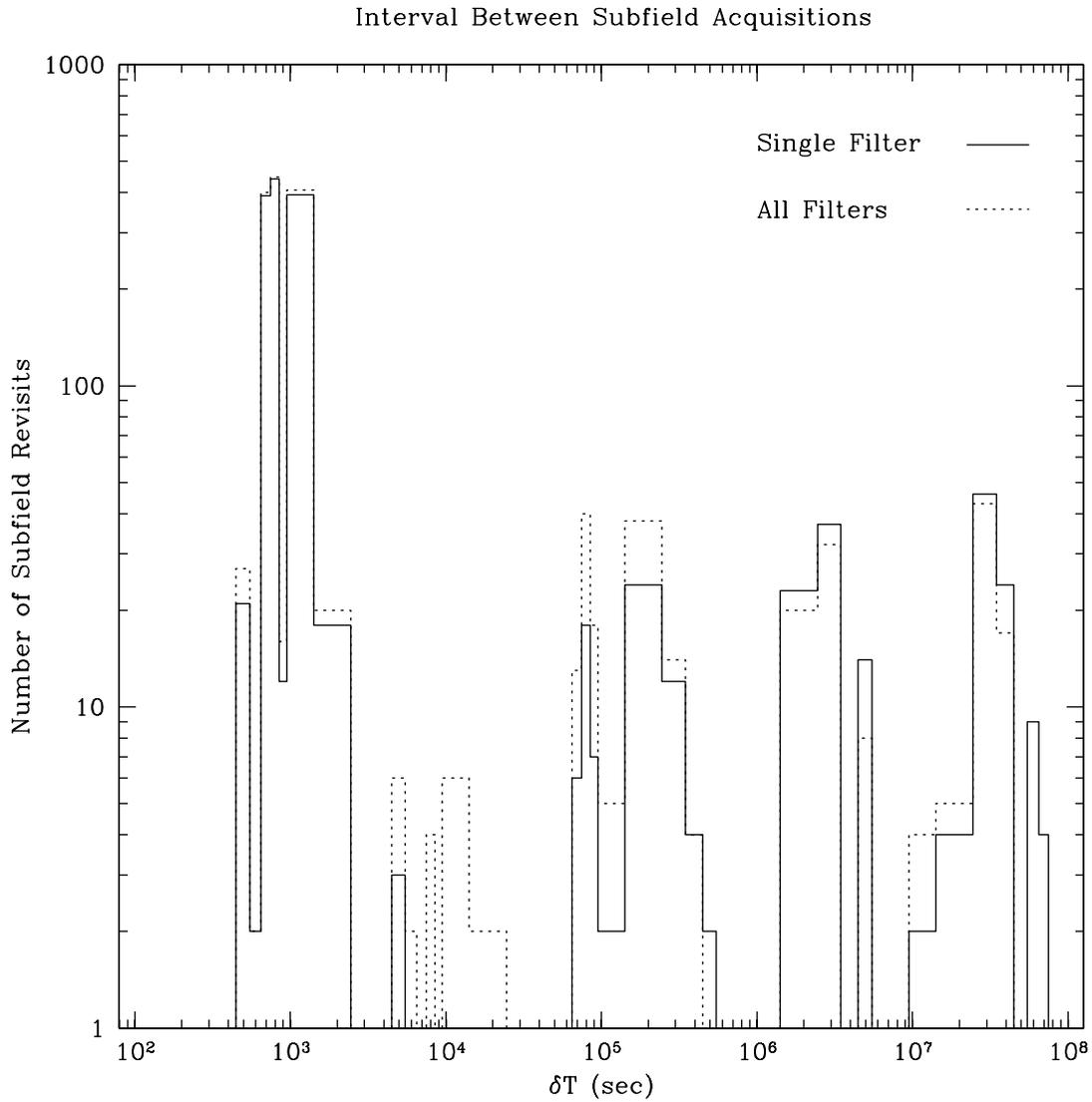}

  \caption{Distribution of sampling intervals between subfield
    observations using DLS transient survey data.  The {\it dotted}
    histogram represents this quantity regardless of the filter of
    observation, whereas the {\it solid} histogram represents the same
    interval restricted to identical subfield--filter combinations.}

\label{fig-samp}
\end{figure*}

The ensemble of Deep Lens Survey images considered for this
publication was taken from 14 observing runs between 2002 November and
2003 April where the transient pipeline was operating efficiently.
Figure~\ref{fig-samp} displays the distribution ({\it dotted} line) of
temporal intervals between subfield reacquisitions, integrated over
all subfields.  The {\it solid} line shows this same quantity, but
only for reacquisitions in the same filter as the previous
observation.  Since variability is detected through comparison of
observations taken in similar filters, the integral under this {\it
solid} subfield--filter histogram up to timescale $\tau$ defines our
overall sensitivity to transients at timescale $\tau$.  The integral
under the {\it dotted} histogram represents our followup capability at
a given $\tau$.

Figure~\ref{fig-samp} indicates our primary sensitivity is to
variability on thousand--second timescales.  In this subset we have a
total integrated exposure time of 10.2 days distributed over 425
$B$--band, 464 $V$--band, and 393 $R$--band images.  At $0.6 \deg
\times 0.6 \deg$ per image, this yields a total exposure $\exposure$
of $3.7$ deg$^2$--days for sensitivity to $10^3$ second events down to
24$^{th}$ magnitude.  We have a factor of 17 smaller $\exposure$ for
inter--night observation intervals at $10^5$ seconds, and also at
inter--month observation intervals at $10^6$ seconds.  A planned
re--analysis of all survey data for transients will allow for a
precise determination of $\exposure$ at all available timescales.

\subsection{Pre--pipeline Image Calibration}

Our typical subfield observation comes as a 5--image dither pattern
controlled using the {\tt IRAF}\footnote{IRAF is distributed by the
National Optical Astronomy Observatories, which are operated by the
Association of Universities for Research in Astronomy, Inc., under
cooperative agreement with the National Science Foundation.} script
{\tt mosdither}.  Standardized information is written in each of the
images' {\tt FITS} headers, and is used in the pipeline to
automatically associate each quintuple of images.

Basic image calibration is done before each image is sent to the
transient pipeline.  The {\tt IRAF} task {\tt mscred.ccdproc} is used
to correct for cross--talk, apply overscan correction and trim the
image, and bias and flat--field correct the images (we generally take
calibration sequences in the afternoon to allow for real--time
reductions).  In the case of $z'$--band data, we need to apply a
fringing correction, requiring an accumulation of images that
generally precludes real--time reductions.  Finally, the images are
registered to the World Coordinate System (WCS) on the sky using {\tt
mscred.msccmatch} with the NOAO:USNO-A2 \citep{usno-v2} catalog.  The
images are then copied to a directory that is continuously searched by
the transient pipeline image reaper.

\subsection{Automatic Pipeline}
\label{sec-pipeline}

Our transient pipeline uses the OPUS environment \citep{opus} as a
backbone sequencing a series of Python and Perl language scripts, each
defining a ``stage'' in our pipeline.  These stages are defined as
follows:

\begin{itemize}

\item {\tt IN} : copy an image via the SCP protocol from the {\tt
IRAF} reduction machine to the transient pipeline machine.

\item {\tt RI} : register and pixel--wise resample an incoming image
to a fiducial template image of the subfield.  WCS information is used
as a starting point for determining the registration coefficients.
The astrometric template is the same for all passbands, and is
constructed from the very first observation of a subfield.

\item {\tt CI} : detect and catalog objects in the registered image
using SExtractor \citep{bertin96}.  For our detection thresholds, we
use SExtractor parameters {\tt DETECT\_THRESH} = {\tt ANALYSIS\_THRESH} =
2.5.

\item {\tt HI} : difference each dither against the first pointing,
using a modified version of the \citet{alard-isis00} algorithm.
Modifications include internal methods to register and remap pixels
based upon WCS information, comprehensive discrimination of
appropriate regions to determine the image convolution kernel, figures
of merit to determine the direction of convolution, robust methods to
ensure complete spatial constraint on the variation of the kernel, and
the propagation of noise through the convolution process

\item {\tt FI} : catalog the difference image and filter object
detections, including the calculation of adaptive second moments
\citep{bj2002}.  These moments, originally designed for accurate shape
measurements in our weak lensing pipeline, are used here to verify the
integrity of each candidate.  Cuts on the results of this analysis
(e.g size criteria for cosmic ray rejection) help to reject a
significant fraction of the noise and false positives, as well as a
fraction of the actual transients present.  This stage yields the
short--timescale transients in each dither sequence, primarily solar
system objects, but also including stellar and unknown sources of
variability.

To algorithmically cull false detections from our object lists, we
first construct exclusion zones around bright stars, using the
NOAO:USNO-A2 \citep{usno-v2} catalog, and along bad CCD columns, using
bad pixel masks supplied with the {\tt mscred} package.  Given a list
of candidates SExtracted from the difference image, we further analyze
each object in both the original input image and the difference image.
In the input image, we require the object's semi--minor axis, as
computed from adaptive second moments, to be greater than 1 pixel
($0.27\arcsec$) as a cut against cosmic rays.  In the difference
image, we further require second moments greater than 0.9 pixels,
uncertainties in the orthogonal ellipticity components less than 0.5
pixels, and integrated flux measurements greater than 0 counts.  We
also examine all pixels within 20 pixels of a candidate, and require
$50\%$ of all pixels to be neither negative nor masked, and $70\%$ of
non--masked pixels to be positive valued.  These additional cuts
safeguard against bad registrations or subtractions, which tend to
leave a number of contiguous pixels above the detection threshold, but
also regions of systematic dipole residuals.  This does impact our
efficiencies for slowly moving objects such as KBOs.  By correctly
propagating noise arrays through the remapping and difference imaging
process, it is possible to compare projected noise properties of the
image to empirical measurements, and automatically identify and reject
such regions.

\end{itemize}

At this point, the individual images are co--added to make a deeper
representation of the subfield and searched for longer timescale
transients :

\begin{itemize}

\item {\tt SI} : co--add the dither sequence of images into a single
deep image of the subfield, providing complete spatial coverage by
filling in MOSAIC chip gaps.

\item {\tt IT} : identify prior templates of the same subfield, in the
same filter, to use as templates for detecting variability on longer
timescales.

\item {\tt HS} : difference the deep image stack constructed in stage
{\tt SI} against the prior templates.

\item {\tt FS} : catalog and filter the deep difference images.
Longer timescale transients are revealed at this stage, primarily
supernova and active galactic nuclei (AGN), and stellar variability.
We employ similar object cuts as in stage {\tt FI}.

\item {\tt CL} : clean up and archive data files, and inform the
observers that the sequence of observations is ready to be reviewed.

\end{itemize}

Our automated cuts typically reduce the number of candidates per image
from several hundreds to several tens.  After the last stage of the
pipeline, the observer must manually categorize the remaining
candidates.  A graphical user
interface displays the sequence of difference images.  Each is
presented as a triplet of postage stamps centered on the candidate and
showing the prior template, current image, and difference between
them.  The nature of the transient is determined using a variety of
factors, which can be divided into temporal and contextual
information.  Temporal discriminants include astrometric motion of the
transient, indicative of solar system origin, as well as the timescale
of photometric variability for stationary objects, which can help
discriminate between populations of variability with known timescales,
or to single out an object varying with unexpected rapidity.
Contextual discriminants include the existence of a precursor or host
object, and the proximity and relative location of the transient
event.

During visual classification of a transient, a set of data files are
automatically generated locally on disk, after cross--checking for
spatial coincidence with previous transients, or proximity to
approximate ephemerides for moving objects.  An ephemeris is generated
for moving objects, and for stationary objects, a finding chart.
These files are copied to our remote web server where a periodic
daemon executes a command to regenerate our web page based upon the
data files present.  Thus, every 15 minutes our web page is brought
into sync with the transients detected at the telescope.  At the
conclusion of each run, all moving objects are reported to the Minor
Planet Center\footnote{\url{http://cfa-www.harvard.edu/iau/mpc.html}}
and the most convincing supernova candidates to the Central Bureau for
Astronomical
Telegrams\footnote{\url{http://cfa-www.harvard.edu/iau/cbat.html}}.

The magnitudes resulting from photometry on difference images are
denoted by the symbol $\mathcal{M}$.  This represents the magnitude of
a static object whose brightness is equal to the measured change in
flux $\Delta f$, $\mathcal{M}\equiv-2.5~{\rm log}~(\Delta f) + m_0$,
where $m_0$ is the photometric zero--point of the template image.
This quantity can be related to the change in magnitude $\Delta M$ of
an object with quiescent magnitude $M$ through the relation
\begin{eqnarray}
\label{eqn-dm}
\Delta M  = -2.5 ~ \log \left( 10^{ (M - \mathcal{M}) \over 2.5} + 1 \right).
\end{eqnarray}
In the limit where the variability comes from a precursor fainter than
the detection limit, such as for supernovae, then $\mathcal{M}$
represents the magnitude of the transient itself.  In the following,
we report all magnitudes derived from difference imaging in units of
$\mathcal{M}$.

We are further pursuing algorithms that might automatically and
optimally discriminate between transient classes, which will be a
necessity for larger scale transient searches.  One such example is
the GENetic Imagery Exploration software package (GENIE:
\mycite{genie}).  GENIE uses genetic algorithms to assemble a sequence
of image operators that optimally extract features from
multi--spectral data, and might be naturally be applied instead to our
multi--temporal sequences.

All archived transients are made available on the internet at
\url{http://dls.bell-labs.com/transients.html}, generally within a few
hours of observations.  Transients are broadly categorized as
``moving'' (slow or fast objects are highlighted), ``variable star'',
``supernova'', ``agn or supernova'' (if this cannot be distinguished
based on the host morphology and transient location), or ``unknown'',
a broad category indicating objects whose astrophysical nature is
uncertain.  This includes very rapidly varying objects, or transients
with no obvious precursor or host.

\section{Astrometrically Variable Objects}
\label{subsec-movers}

Roughly 75\% of the transients detected in the DLS are astrometrically
variable solar system objects.  Four of the five DLS fields are within
15$^\circ$ of the ecliptic, and in these fields we generally find of
order ten moving objects per subfield visit.  The total number of
moving objects detected through spring 2003 is 3651, with 900 detected
in $B$, 1508 detected in $V$, and 1487 detected in $R$ (some were
detected in multiple filters).  Our detection efficiency for moving
objects has not been modeled.

Because of our observing pattern, most of the moving objects are
observed on a single night and never recovered.  A main belt asteroid
moving at $\sim$30\arcs\ hr$^{-1}$ near opposition leaves a subfield
within a few days.  As a result, only about one-third of our moving
objects have been observed on multiple nights, and 333 of these have
received Minor Planet Center provisional designations, which are
listed on our website.  Similarly, we are able to measure colors for
only a fairly small fraction of these objects.  Therefore, we do not
attempt a thorough analysis of orbital parameters, albedos, and sizes.
Rather, we give an overview of the dataset and then detail a few
especially interesting objects.

\begin{figure*}[T]
  \plotone{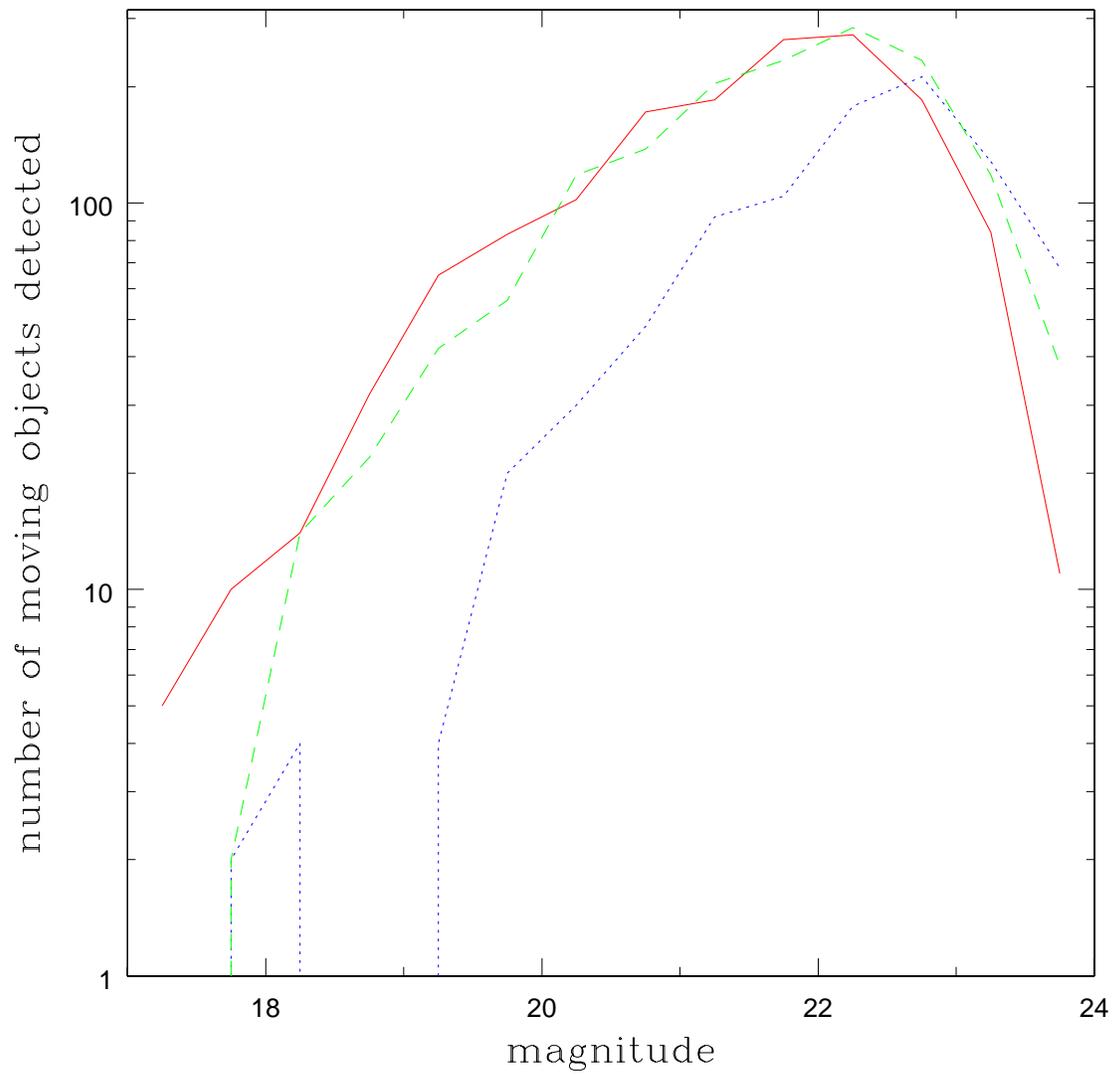}

  \caption{Magnitude distribution of moving objects detected in the
    DLS. Detections in the $B$, $V$, and $R$ passbands are represented
    by dotted, dashed, and solid lines, respectively. Some objects
    were detected in more than one filter, but most were not. }

  \label{fig-astmag}
\end{figure*}

Figure~\ref{fig-astmag} shows the magnitude distribution of all moving
objects discovered, separately for $B$, $V$, and $R$.  Note that the
magnitude range is displaced from the magnitude range in which our
efficiency for stationary point sources is nonzero
(Figure~\ref{fig-eff}).  This is because the vast majority of moving
objects are significantly trailed during our 600 or 900 second
exposures.  An asteroid at the bright end can therefore be much
brighter than a point source before saturation sets in, and at the
faint end must also be brighter to rise above the detection threshold.

\begin{figure*}[T]
  \plotone{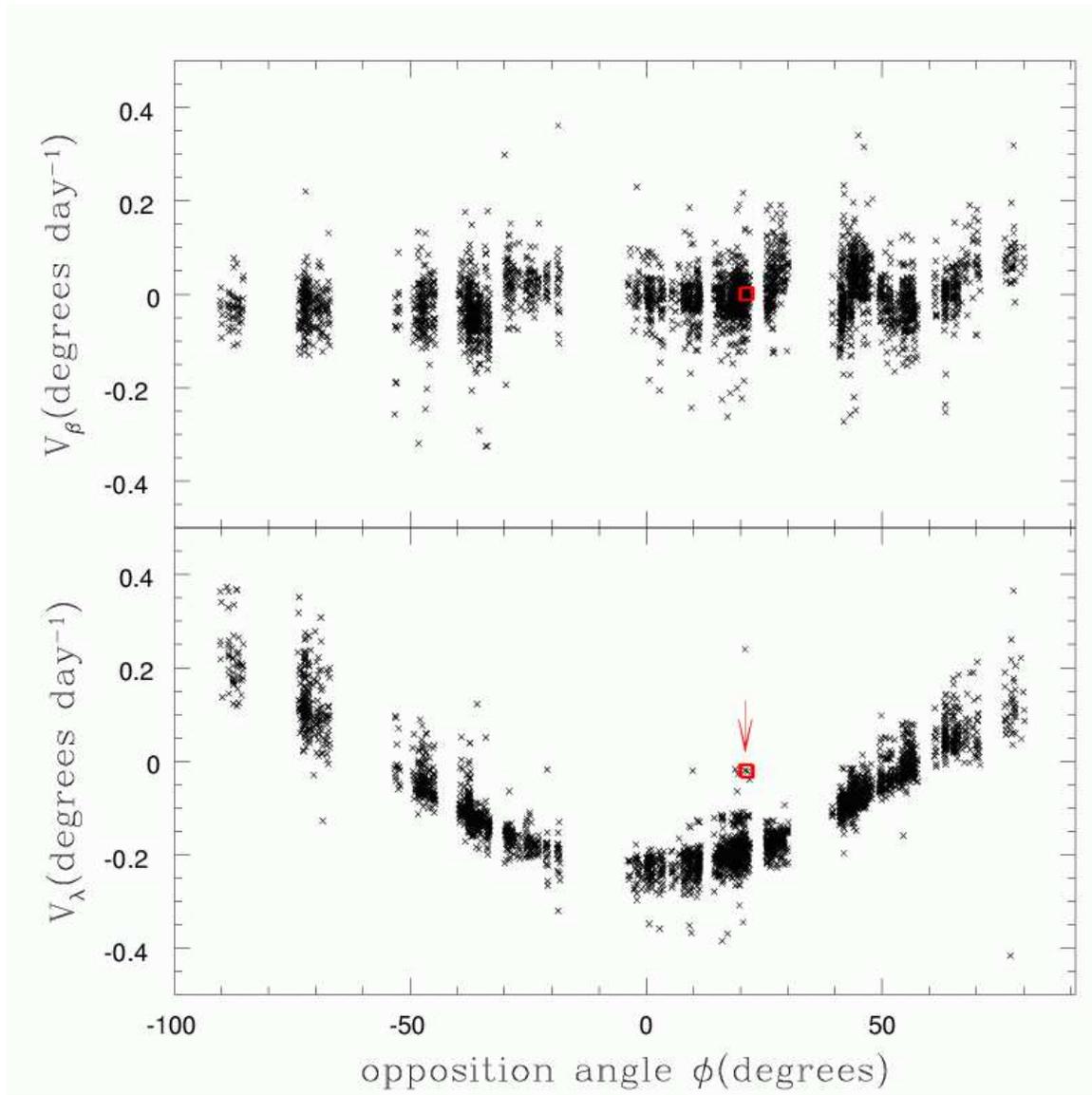}

  \caption{The moving object velocity distributions in ecliptic
    coordinates.  Two Kuiper Belt objects detailed in
    Section~\ref{sec-kbo} are marked with a red box.  They clearly stand
    out from the main belt $v_\lambda$ distribution, and neighboring
    points are also likely KBOs.  Overall, KBOs do not stand out in
    the $v_\beta$ distribution.}

  \label{fig-veldist1}
\end{figure*}

Figure~\ref{fig-veldist1} shows the moving objects' velocity
distribution in ecliptic coordinates, highlighting a pair of Kuiper
Belt objects (KBOs) detailed below.  Figure~\ref{fig-veldist2}
provides tentative classifications into asteroid families (main belt,
Trojans, Centaurs, NEOs, etc.) based upon ecliptic velocities,
adopting the boundaries of \cite{ivezic-ast}.

\subsection{Kuiper Belt Objects}
\label{sec-kbo}

The 2003 April CTIO run yielded two Kuiper Belt Objects (KBOs).  Our
pipeline does not include an orbit calculator, so KBOs are not
automatically flagged.  These two examples were selected manually on
the basis of their small ($<3$\arcs\ hr$^{-1}$) angular motions near
opposition.

Transient 207 from the 2003 April CTIO run was discovered on 2003
April 1 (UT), moving at 2.9\arcs\ hr$^{-1}$ near opposition, with a
median $R$ magnitude of 21.5.  Using the method of \cite{bern-kbofit},
we find a semimajor axis, $a$, of 41.0 AU, eccentricity $e$ of 0.022,
and inclination $i$ of 20.1$^\circ$, making this a classical KBO.
Assuming an albedo of 0.1, the diameter of this object is $\sim$475
km, making it one of the larger KBOs known.  Transient 587 from the
same run was discovered on 2003 March 31 (UT) at a $B$ magnitude of
23.3, moving at 2.6\arcs\ hr$^{-1}$, also near opposition.  Our orbit
calculation yields $a = 42.8$, $e = 0.178$, and $i = 1.1^\circ$,
making it a likely scattered KBO.  For an albedo of 0.1, this object
is $\sim$410 km in diameter.  Our observations provide orbital arcs of
5 and 4 days for these KBOs, respectively.

\subsection{Near-Earth Objects}
\label{sec-neo}

\begin{figure*}[T]
  \plotone{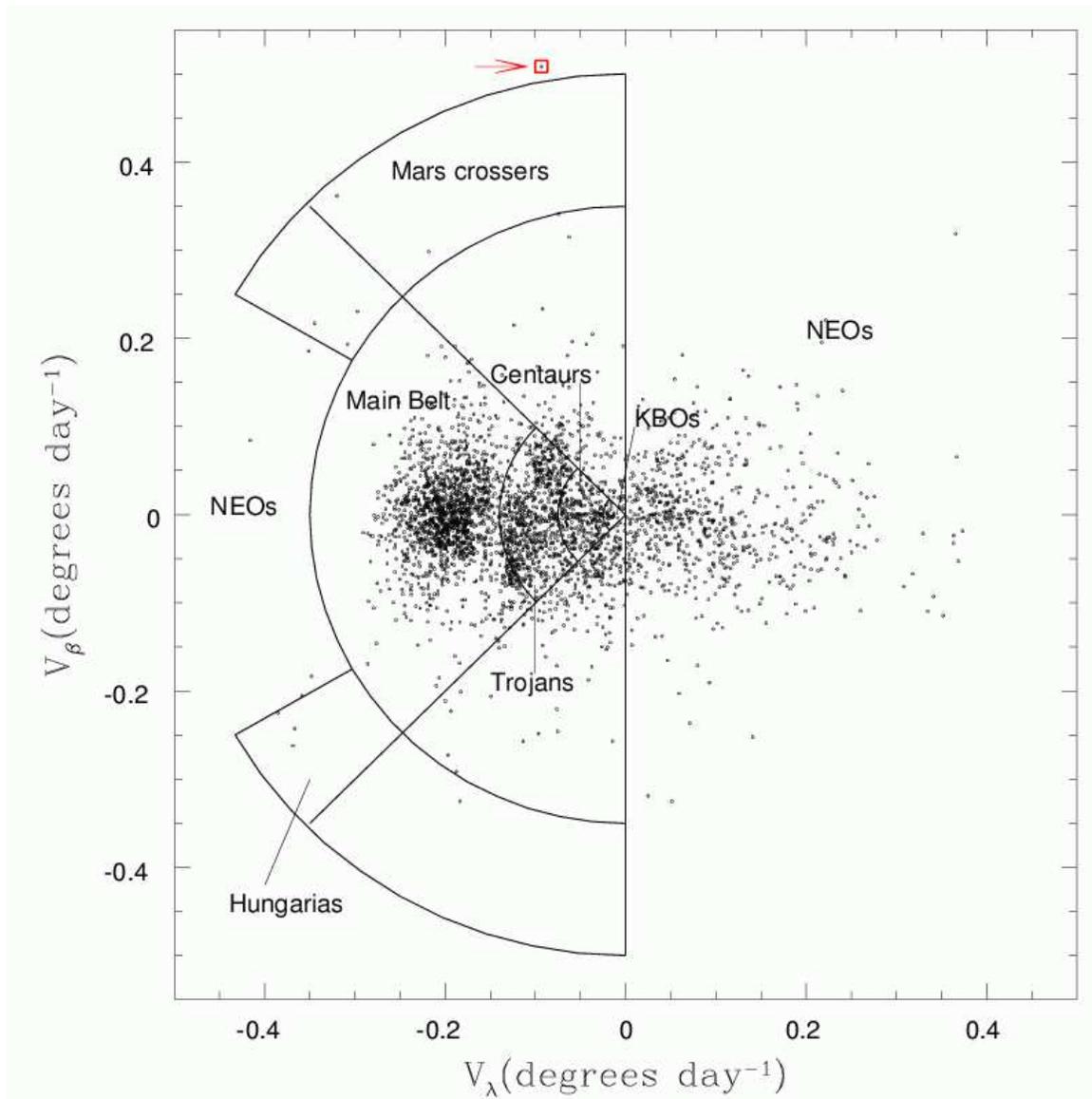}

  \caption{Tentative classification of moving objects into asteroid
    families, based on velocities. The object boxed at top is the
    fast-moving object detailed in Section~\ref{sec-neo}.}

  \label{fig-veldist2}
\end{figure*}

The detection efficiency for fast movers declines with velocity, as
their reflected light is trailed over larger area.  The DLS, with its
long exposures, is therefore far from optimal for detecting Near-Earth
Objects (NEOs).  Nevertheless, we have detected a sizable sample, as
shown in Figure~\ref{fig-veldist2}.  Here we highlight one that was
moving with an unusual velocity, Transient 1363 from the 2003 March
KPNO run.  This object is the fastest solar system object seen during
the transient search, moving at 80\arcs\ hr$^{-1}$.  The box near the
top of Figure~\ref{fig-veldist2} indicates this object's remote
position in the ecliptic velocity diagram.

\section{Photometrically Variable Objects}
\label{subsec-shakers}

We focus here on those transients that have been noted to flare on
thousand second timescales and are detected in at least two images, as
well as on longer--term variability with no detectable originating
host or precursor.  However, our transient catalog also includes a
selection of several hundred variable stars, as well as over 100
supernova candidates, 18 of which pass the requirements for
recognition as defined by the Central Bureau for Astronomical
Telegrams.  These include : 2000bj (\mycite{2000IAUC.7398....2K}),
2000fq (\mycite{2000IAUC.7551....1W}), 2002aj, 2002ak, 2002al, 2002am
(\mycite{2002IAUC.7804....1B}), 2002ax, 2002ay, 2002az, 2002ba,
2002bb, 2002bc, 2002bd, 2002be (\mycite{2002IAUC.7833....1B}), 2003bx,
2003by, 2003bz, 2003ca (\mycite{2003IAUC.8093....1B}).

\subsection{Image Calibration}

We have re--run the difference imaging algorithm for this paper, using
2k x 2k ($8.6\arcmin$ x $8.6\arcmin$) arrays centered on the selected
optical transients (OTs).  We have generally been able to construct
deeper, higher signal--to--noise (S/N) template images than those
available to us at the time of detection.  For photometric
calibration, we need only calibrate the zero points of these template
images, as the difference imaging convolution normalizes each input
image to the template's photometric scale.  The $B$, $V$, and $R$
templates are normalized to the \cite{landolt-92} system, with
additional errors (typically $5\%$) added in quadrature to account for
instrumental differences between the KPNO, CTIO, and \cite{landolt-92}
systems.  All $z'$ templates are initially calibrated to the
\cite{sloan-cal} $AB$ system, and presented here in the Vega system by
adopting an offset of $0.54$ magnitudes ($z'_{Vega} = z'_{AB} -
0.54$).  The transformation between photometric systems depends on the
instrumental filter response and on Vega's spectral energy
distribution, adding an additional $0.08$ magnitude uncertainty to our
$z' = z'_{Vega}$ measurements.

The template brightnesses of the precursor, or host, objects were
determined using {\tt MAG\_BEST} from SExtractor \citep{bertin96},
except in the case of OT 20010326 (Section~\ref{sec-trans52} below).
For this precursor, which appears unresolved, we first used the {\tt
IRAF} {\tt noao.digiphot.daophot.psf} package to determine the local
PSF from nearby isolated stars, and subtracted these stars and the
transient precursor from the image using {\tt
noao.digiphot.daophot.allstar}.  

For those template images where we are not able to detect the OT
precursor/host, we define the point source detection limit in
accordance with the NOAO Archive definition of photometric depth
(T. Lauer, private communication), as
\begin{eqnarray}
\label{eqn-limit}
m_l = m_0 - 2.5 ~ {\rm log} (1.2 ~ W ~ \sigma_{\rm sky} ~ n),
\end{eqnarray}
where $m_0$ is the magnitude of one ADU, $W$ is the seeing in pixels,
$\sigma_{\rm sky}$ is the dispersion of the image around its modal sky
value, and we choose $n = 3$ $\sigma$ as our detection limit.  We
determine detection limits in the difference images by measuring
directly our ability to recover input stellar PSFs.

For the difference imaging, we choose the best--fit convolution kernel
that varies spatially to order one across our 2k x 2k subimage, and a
sky background with a similar degree of variation.  We fit these
spatial variations over a 20 x 20 grid, using a convolution kernel
with a half--width of 10 pixels.  The function of the convolution
kernel is to degrade the higher quality image to match the PSF of the
lesser quality image.  Objects are detected in each grid element to
constrain locally the convolution kernel through a pixel--by--pixel
comparison between images.  A global fit is next performed, using each
element as a constraint on spatial variation of the kernel.  Several
sigma--clipping iterations ensure that sources of variability, such as
variable stars or asteroids, are not used as constraints for this
kernel.  If such an element is rejected from the global fit, a
replacement object within this element is chosen to ensure maximal
spatial constraint.  Application of the final convolution kernel to
the entire image and pixel--by--pixel subtraction yields the
difference between the two images, with a resulting point spread
function of the lesser quality image, and photometric normalization to
the template image.  The transient fluxes were determined using the
{\tt IRAF} {\tt noao.digiphot.apphot.phot} package.
The relative transient magnitudes are typically determined to better
than $\sim 4\%$, with the remaining uncertainty due to absolute
calibration of the template zero--point.

\subsection{Optical Transients Without Hosts}
\label{subsec-OTnohost}

We have detected several optical transients with no obvious hosts, and
which are seen to vary over the timescale of months
(e.g. \mycite{2003IAUC.8093....2B}), similar to supernovae and GRB
afterglows.  One of the more luminous examples of this class of
objects is OT 20020112 \citep{2002IAUC.7803....2B}, originally
reported as Transient 139 from our 2002 January run
\footnote{\url{http://dls.bell-labs.com/transients/Jan-2002/trans139.html}}.
Information on this event is summarized in Table~\ref{tab-short},
including the DLS subfield in which it was detected (F4p13), and
position on the sky in J2000 coordinates.  We include limits on a
precursor or host for this event, which is not detected to deeper than
$27^{th}$ magnitudes in the $B$, $V$, and $R$ passbands, and $\sim
25^{th}$ magnitude in $z'$.  Table~\ref{tab-short} also lists limits
on the precursor/host brightness after taking into account Galactic
reddening, assuming a $R_V = 3.1$ extinction curve \citep{schlegel}.

\begin{figure*}[t]
  \plotone{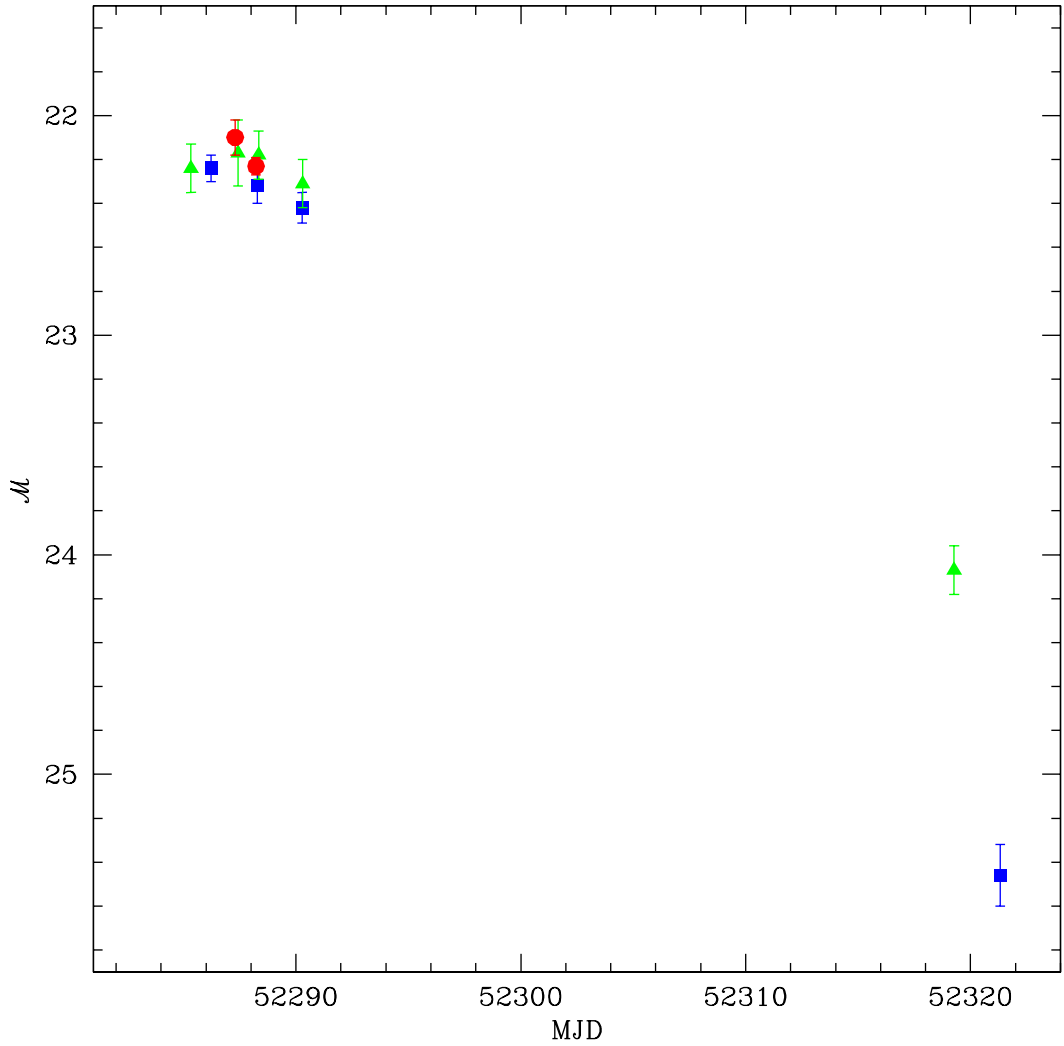}

  \caption{Lightcurve OT 20020112 in differential flux magnitudes
  $\mathcal{M}$ in the $B$ (square datapoints), $V$ (triangle), and
  $R$ (circle) passbands.  The data were obtained at the CTIO 4--m
  Blanco telescope and the KPNO 1.3m MDM telescope.  All data from a
  given MJD are averaged together after being placed on similar
  magnitude systems during the difference imaging process.  Lightcurve
  information is presented in Table~\ref{tab-phot}. There is no
  apparent host for this variability in images obtained prior to and
  subsequent to the observed event.}

\label{fig-ot20020112}
\end{figure*}

OT 20020112 was detected in observations taken 2002 January 12.
Previous subfield observations were taken 2001 March 27, so the age of
the transient at detection is poorly constrained.  Information on this
transient was quickly posted to the Variable Star NETwork
(VSNET\footnote{\url{http://vsnet.kusastro.kyoto-u.ac.jp/vsnet/index.html}}),
prompting a sequence of followup observations from the 1.3m
McGraw--Hill telescope (MDM).  The lightcurve for this transient is
shown in Figure~\ref{fig-ot20020112}, and represents a daily averaged
lightcurve of all observations from the DLS and MDM.  The event was
re--imaged 35 days after detection as part of DLS survey observations,
and had faded and reddened significantly.  Detailed photometry for
this event is listed in Table~\ref{tab-phot}.  The magnitudes listed
are in differential flux magnitudes $\mathcal{M}$ -- however, given
the non--detection of a precursor or host, this also represents the
traditional magnitude $M$ of this transient.

\begin{figure*}[t]
  \includegraphics[width=0.55\columnwidth,angle=-90]{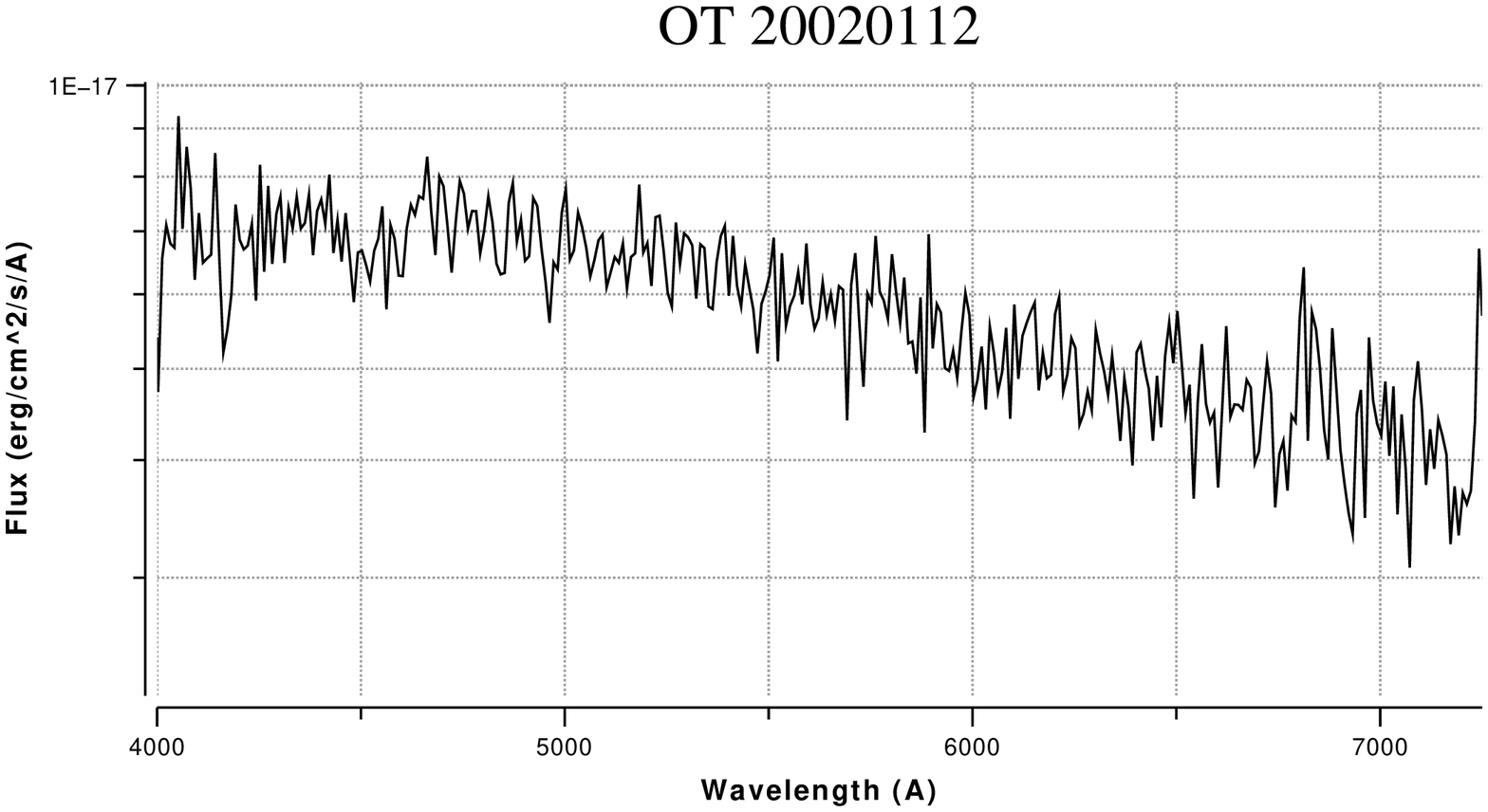}
  \caption{Spectrum of OT 20020112 obtained 6 days after detection.
    The phase of the event is uncertain.  Weak emission features
    indicate a possible redshift of $z = 0.038$.  However, we are
    unable to detect a host to $R > 27.6$.  This transient represents
    a class of objects that vary on timescales of tens of days, much
    like supernovae or potentially optical GRB orphans, but have no
    apparent host galaxy.}

\label{fig-spec20020112}
\end{figure*}

OT 20020112 was also observed spectroscopically with the Las Campanas
Observatory 6.5-m Baade telescope (+ dual imager/spectrograph LDSS2)
on 2002 January 18.  The spectrum in Figure~\ref{fig-spec20020112} is
characterized by a blue continuum with no obvious broad features, and
marginal evidence for emission lines H$\alpha$ at $z = 0.038$ and
[NII] 6583 at $z = 0.039$.  The continuum features and lack of short
timescale variability are consistent with a Type II SN caught close to
maximum brightness.  However at $z = 0.038$, the observed brightness
($V \sim B \sim 22.1$) is $3$ magnitudes dimmer than a typical Type II
SN near maximum.  The blue continuum makes a reddened SN explanation
unlikely.  Examination of the FIRST 1.4 GHz radio catalog
\citep{first-97} yields a limit of 0.94 mJy/beam at this position,
with no object closer than $2.4\arcmin$.  A search of the ROSAT source
catalog \citep{rosat-faintcat}, and a more comprehensive search of
archival X--ray and gamma--ray data through HEASARC\footnote{A service
of the Laboratory for High Energy Astrophysics (LHEA) at NASA/ GSFC
and the High Energy Astrophysics Division of the Smithsonian
Astrophysical Observatory (SAO),
\url{http://heasarc.gsfc.nasa.gov/db-perl/W3Browse/w3browse.pl}}
yields no archival sources within $10 \arcmin$.

Thus the nature of this object remains unknown, a situation compounded
by the lack of an obvious host galaxy to $R > 27.6$.  We note the
observed $B - V$ color change during the event is inconsistent with a
spectral energy distribution of the form $F_\nu \propto \nu^{\beta}$,
such as that expected for GRB afterglows (e.g. \mycite{sari-spec},
\mycite{nakar-2002}).  However, evolution of the index $\beta$ could
lead to the observed behavior.  Future OT searches will require
facilities dedicated to followup of such events, to associate or
distinguish them from the expected supernova and GRB afterglow
populations.

\subsection{Rapid Variability Optical Transients}
\label{subsec-unknown}

Over the course of the survey we have also detected three optical
transients on thousand--second timescales that were present in
multiple images, and whose precursors were not immediately recognized
as stellar in all available passbands.  In all cases, we were able to
identify a precursor or host object after the fact.  These are
designated OT 20010326, OT 20020115, and OT 20030305.
Table~\ref{tab-short} lists the observed characteristics of the
precursors or hosts for these transients in their quiescent state.
Table~\ref{tab-short} also includes magnitudes corrected for Galactic
reddening for OT 20010326 and OT 20030305, in the case that they lie
outside the Galactic dust layer.

When putting our subsequent results in a cosmological context, we
assume a WMAP cosmology of $H_0 = 71, \Omega_M = 0.27, \Omega_\Lambda
= 0.73$ \citep{wmap-results03}.

\subsubsection{OT 20010326 : Mar 2001, Transient 52}
\label{sec-trans52}

OT 20010326 was detected 2001 March 26.2 as the $52^{nd}$ transient
during that particular run
\footnote{\url{http://dls.bell-labs.com/transients/Mar-2001/trans52.html}}.
Photometry for this event lightcurve, including the constraining
observations immediately preceding and following the event, is listed
in Table~\ref{tab-phot}.  After verification, an alert was dispersed
via e--mail on Mar 26.4 requesting immediate followup observations.

\begin{figure*}[t]
  \plottwo{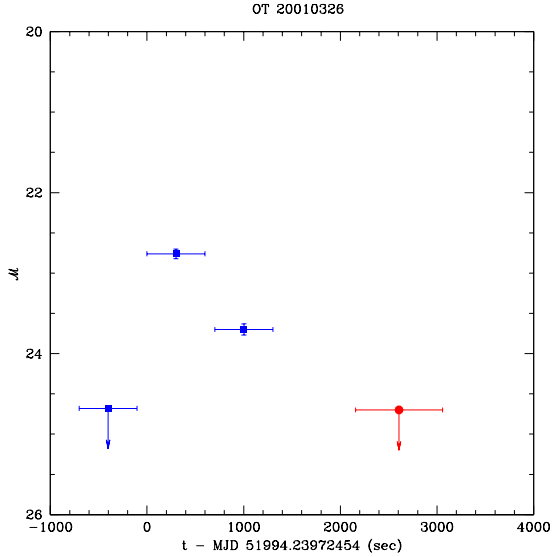}{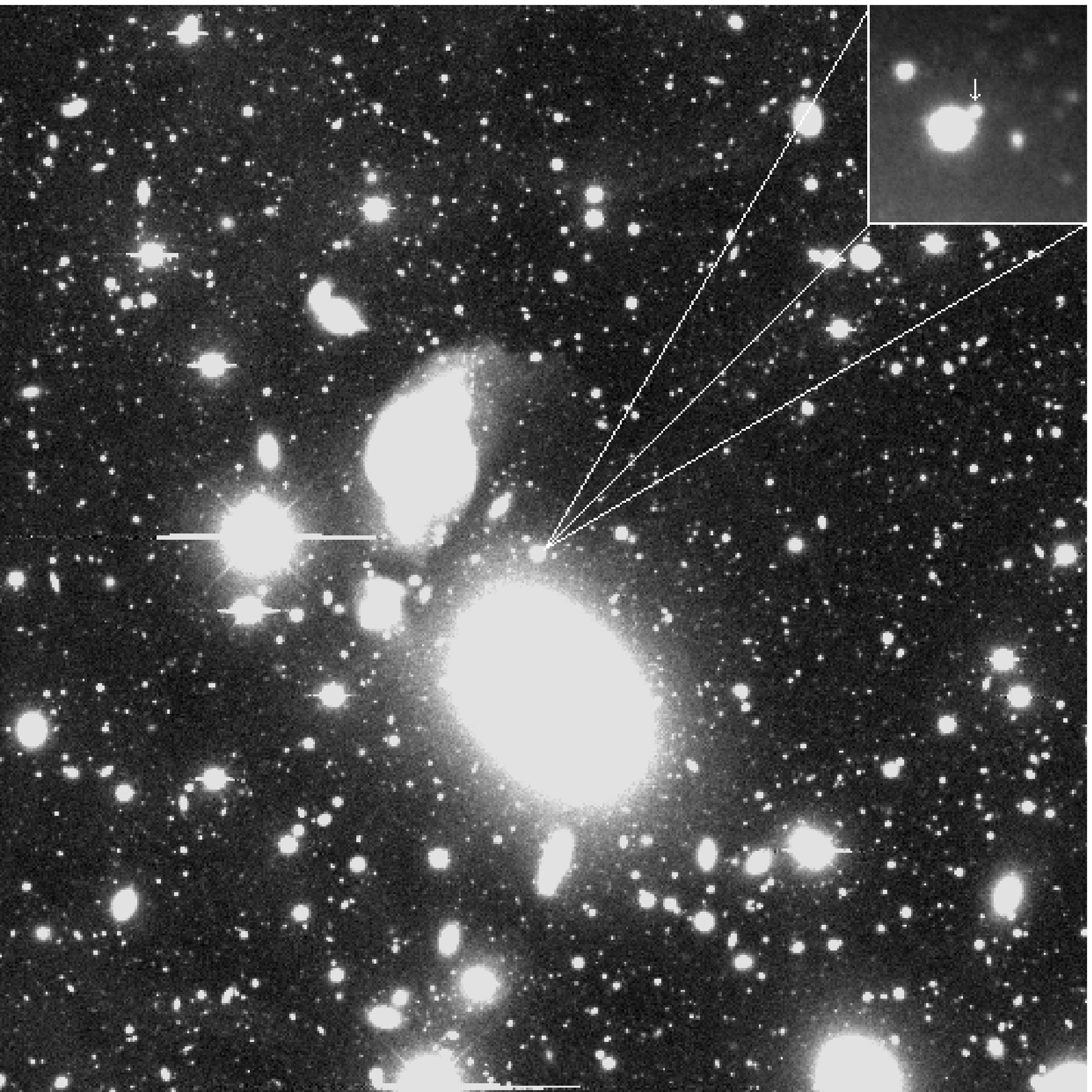}

  \caption{$B$ and $R$--band lightcurve (square and circular symbols,
    respectively) in differential flux magnitudes $\mathcal{M}$ (see
    Equation~\ref{eqn-dm}), and $8.6\arcmin$ x $8.6\arcmin$ $R$--band
    template image, of OT 20010326.  The latter shows its position in
    the field of Abell 1836, at $z = 0.036$.  The precursor is to the
    immediate north--west of the (saturated) brighter star, where
    north is up and east is to the left.  Radio source and elliptical
    galaxy PKS 1358-11 is to the south, and spiral galaxy LCRS
    B135905.8-112006 is to the north--east.}

\label{fig-trans52}
\end{figure*}

The event lightcurve is shown in Figure~\ref{fig-trans52}, and is
characterized by a null detection in $B$, followed by a $B$--band peak
and immediate fall--off.  Given the range of allowed turn--on times,
from $100s$ before the start of the detection observation (to account
for read--out time between images) to $1s$ before the shutter closes
in the detection image, we can limit a power--law index for the flux
decay $f \propto t^{-\alpha}$ of $0.80 < \alpha < 1.2$.

Radio observations of this OT were undertaken at the VLA on 2001 March
30.  Limits on radio flux at 8.5 GHz are $-0.1 \pm 0.3$ mJy.
Subsequent observations of the event revealed a host galaxy or
precursor object visible in the $V$, $R$, and $z'$ passbands, and a
limit of $B > 26.4$.

This event occurred in the field of galaxy cluster Abell 1836, at a
redshift of $z = 0.037$.  Figure~\ref{fig-trans52} also shows an
$8.6\arcmin$ x $8.6\arcmin$ $R$--band template image centered on the
transient event.  The precursor/host is displayed at higher resolution
in the upper--right corner as the dim object $3.5 \arcsec$ north--west
of the brighter star.  A single archival HST image of cluster galaxy
PKS 1358-11 also covers the location of the observed transient.  This
$500s$ integration in the F606W filter was obtained Jan 25, 1995 as
part of a nearby AGN survey by \cite{malkan-agn}.  The precursor
appears unresolved in this image.  We compare this precursor position
with the location of two background galaxies, and over a baseline of
7.0 years find proper motion limits of $0.004 \pm 0.004 \arcsec {\rm
yr}^{-1}$.

It is particularly difficult to photometer and calibrate this OT
precursor/host, as it occurs $3.5\arcsec$ from a bright star that is
saturated, or nearly so, in all of our images.  There is also a very
strong background gradient from the nearby elliptical galaxy PKS
1358-11 and spiral galaxy LCRS B135905.8-112006.  If the OT host is a
member of Abell 1836, it lies only 49 kpc in projection from the core
of PKS 1358-11, and 53 kpc in projection from the center of LCRS
B135905.8-112006.  The HST resolution limit at the distance of Abell
1836 is around 70 parsecs, which does not immediately preclude a
globular cluster host.  However, at a distance modulus of 36.0, the
host absolute magnitude of $M_V = -11.7$ would be considerably
brighter than the globular cluster systems in our own Galaxy
\citep{vdb-gc}.  In addition, the observed $V - R = 1.1$ color of the
object makes it unlikely to be an unresolved globular cluster system
(K.A.G. Olsen, private communication).

\subsubsection{OT 20020115 : Jan 2002, Transient 337}
\label{subsec-ot20020115}

\begin{figure*}[t]
  \plotone{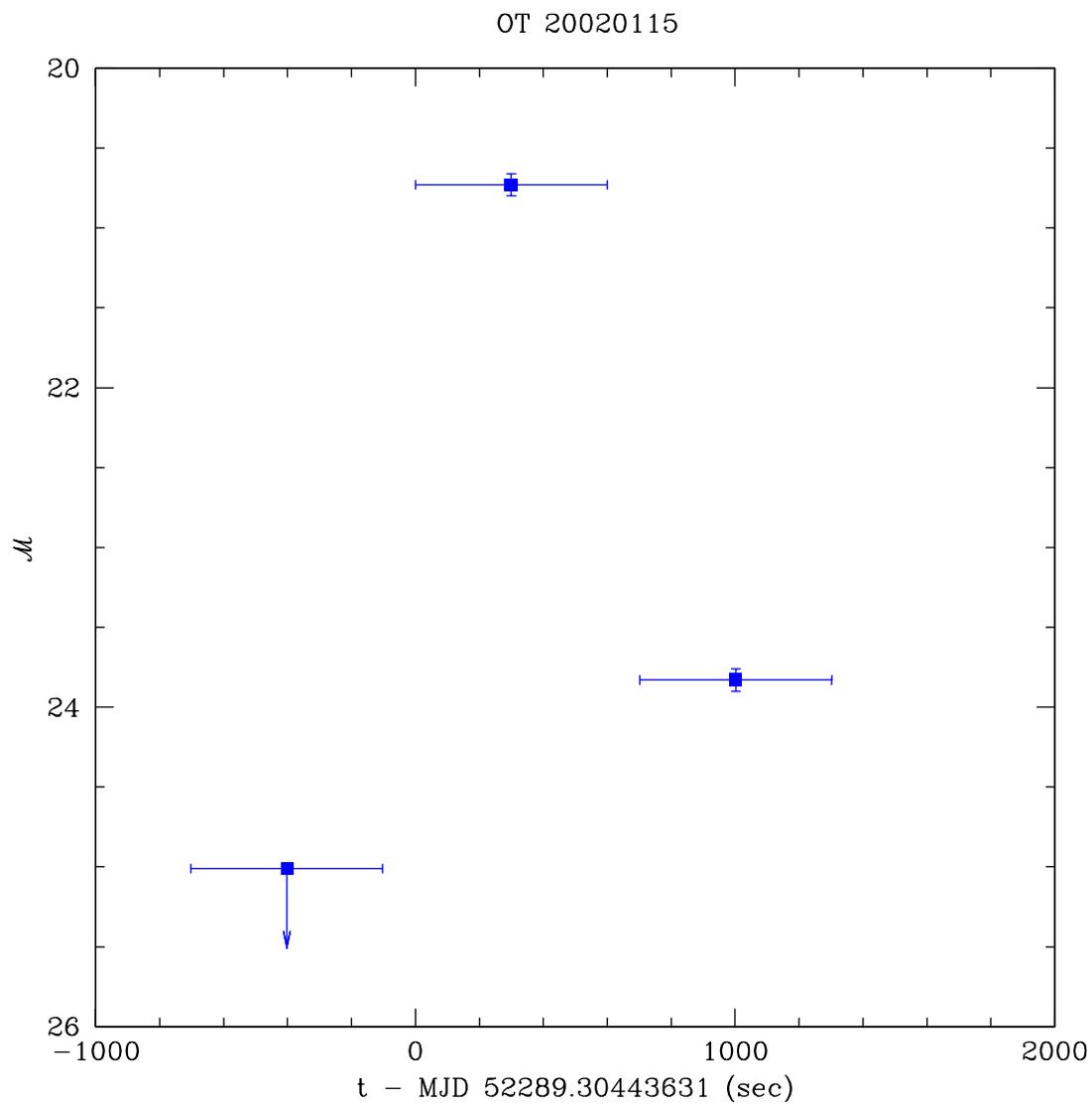}

  \caption{$B$--band temporal lightcurve of OT 20020115 in differential flux
    magnitudes $\mathcal{M}$.}

\label{fig-trans337}
\end{figure*}

Our second fast optical transient, OT 20020115, was initially reported
as Transient 337 from the run ongoing 2002 January
15.3\footnote{\url{http://dls.bell-labs.com/transients/Jan-2002/trans337.html}}.
The lightcurve, whose data are presented in Table~\ref{tab-phot} and
displayed in Figure~\ref{fig-trans337}, is very similar to that of OT
20010326, including $B$--band detection, but it reaches a brighter
$\mathcal{M}_B$, and decays much more rapidly.  The flux decay
power--law index $\alpha$ is constrained to $1.5 \leq \alpha \leq
2.4$.

\begin{figure*}[t]
  \plotone{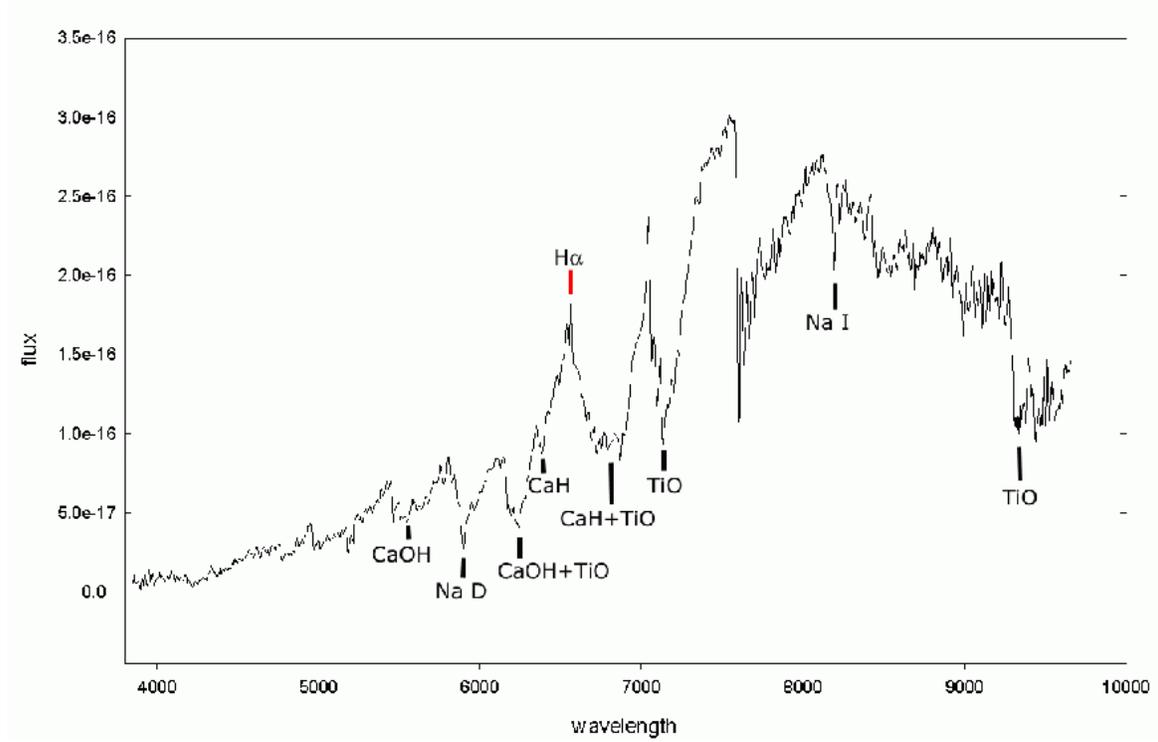}

  \caption{Post--event spectrum of OT 20020115.  The spectrum was
  obtained with the Magellan 1 telescope 3 days after the transient
  event depicted in Figure~\ref{fig-trans337}.  The clear emission in
  H$\alpha$ and absorption by TiO bands indicates Galactic origin.
  Overall, the spectral features and composite colors are consistent
  with a late type (dM4) flare star.}

\label{fig-trans337spec}
\end{figure*}

This event was quickly reported as a potential GRB optical counterpart
to the GRB Circular Network (GCN, \mycite{2002GCN..1217....1B}).  We
obtained spectra of the source soon thereafter --
Figure~\ref{fig-trans337spec} shows 3 x 10 minutes of exposure on the
source with the Magellan 1 telescope + LDSS2 dual image/grism
spectrograph, taken 2002 January 18.3.  The appearance of
zero-redshift emission and absorption features in these spectra are
evidence for Galactic origin, and the GCN alert was amended
(\mycite{2002GCN..1218....1C}).

Subsequent observations in $R$ and $z'$ revealed a very bright, red
precursor object, consistently saturated in $R$--band images.  We also
recover the precursor object in the $B$ and $V$ bands.  With a
quiescent magnitude of $B = 21.3$ and measured peak $\mathcal{M}_B =
20.7$, Equation~\ref{eqn-dm} indicates the precursor varied by more
than 1 magnitude, averaged over the duration of our $600s$ exposure.

The analysis of the photometric observations and object spectrum
(Figure~\ref{fig-trans337spec}) shows OT 20020115 has the colors and
spectrum of a dM4 post--UV Ceti type flare star.  The spectrum shows
the clear presence of late--type M-dwarf spectral indicators:
molecular bands of TiO, CaOH (5530-60\AA~and centered at 6250\AA),
plus CaH at 6385\AA, in addition to weak H$\alpha$ and H$\beta$ in
emission.  With a $V$ magnitude of $19.67 \pm 0.05$, and colors $B - V
= 1.58 \pm 0.09$ and $V - z' = 3.39 \pm 0.10$, these data are
consistent with an M dwarf of spectral type dM4, i.e. a late type
flare star in the quiescent state.  Assuming a typical $M_V = 12$, OT
20020115 would lie at a distance of approximately 350 pc, and 250 pc
above the Galactic plane, well within the scale height for disk
population M dwarfs.

\subsubsection{OT 20030305 : Mar 2003, Transient 153}

\begin{figure*}[t]
  \plottwo{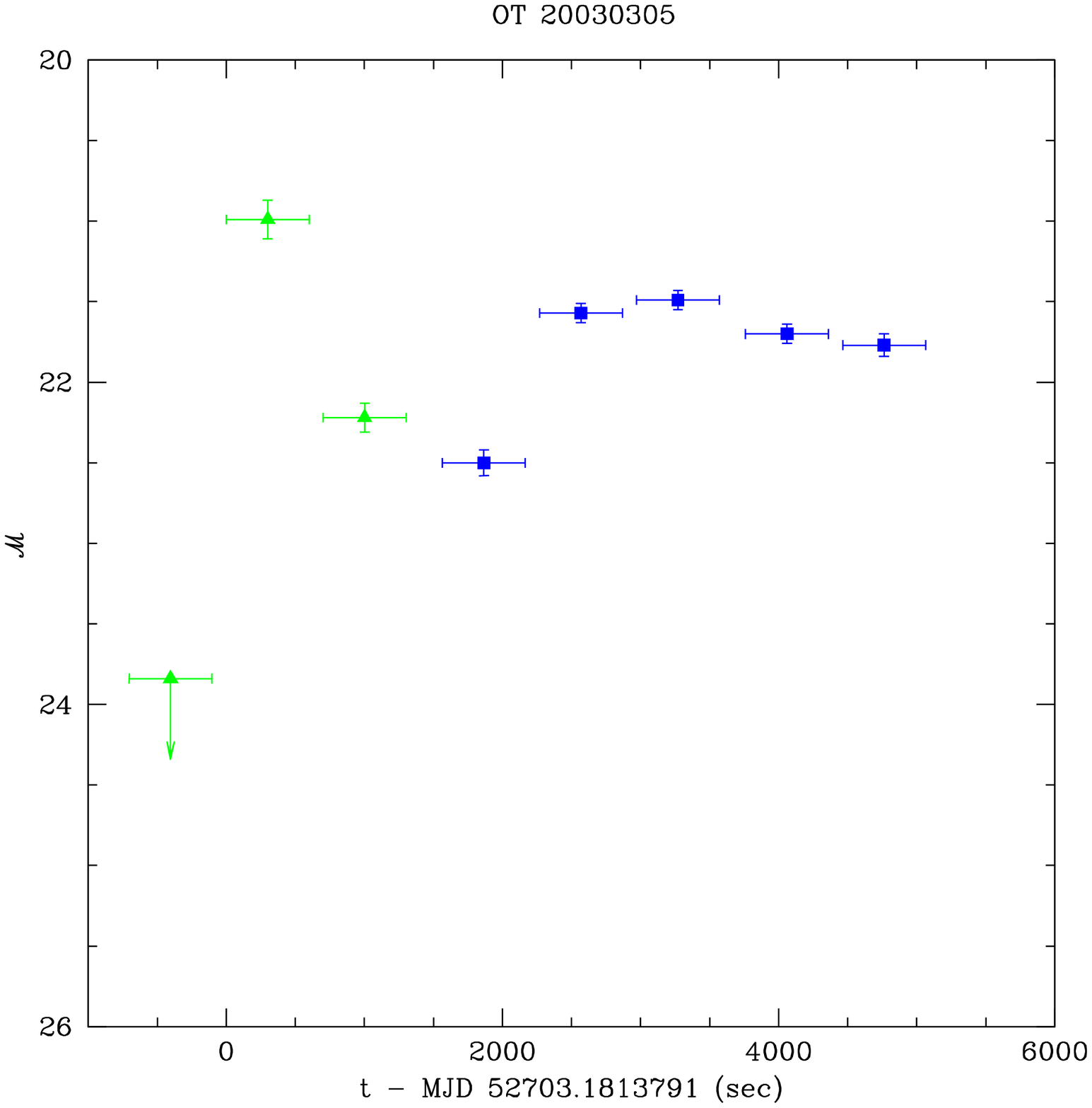}{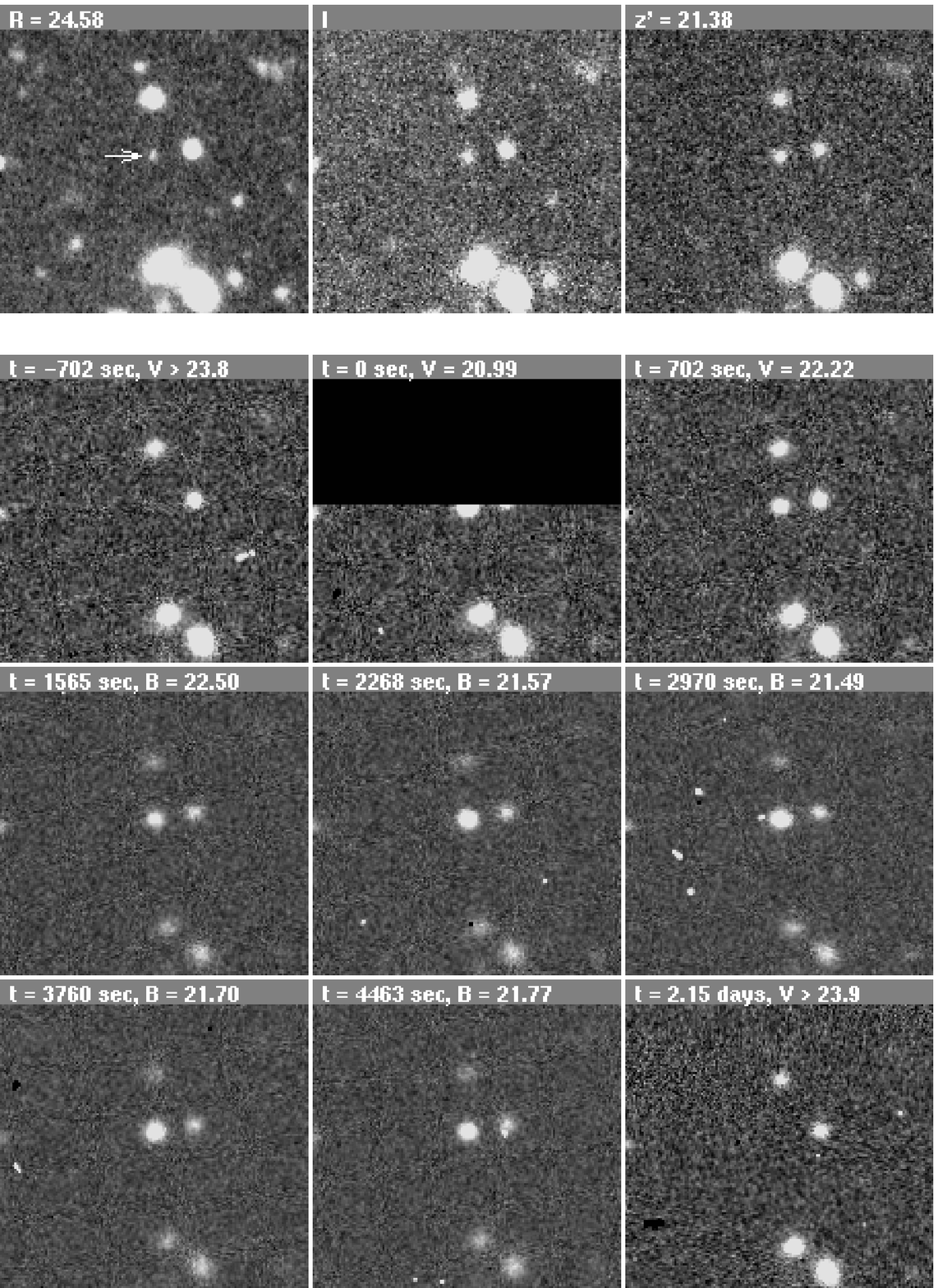}

  \caption{Lightcurve of OT 20030305, in differential flux magnitudes
    $\mathcal{M}$, in the $B$ and $V$ passbands (square and triangular
    datapoints, respectively).  The image sequence corresponding to
    these data points is shown in $40\arcsec$ x $40\arcsec$ panels to
    the right.  The top 3 images show the host, or precursor, of the
    transient in $R$, $I$, and $z'$ (we have no calibration data for
    the $I$--band image).  The images below show the time evolution of
    the event, from the constraining image $700s$ before, to the first
    image 2 days after where it was undetected in $V$.  The time of
    each observation relative to the detection of the event is
    indicated, as well as the measured brightness of the transient,
    where these numbers were derived from photometry on the difference
    images.  Our first detection of this transient landed on the
    boundary of a MOSAIC chip, the dark region in the above detection
    image.  In all images, north is up and east is to the left.  }

\label{fig-trans153}
\end{figure*}

The most dynamic transient, OT 20030305, was detected as Transient 153
from our 2003 March
run\footnote{\url{http://dls.bell-labs.com/transients/Mar-2003/trans153.html}}.
This is the only event detected in the $V$--band, and we immediately
acquired a sequence of $B$--band observations after the $V$--band
series.  The event lightcurve is shown in Figure~\ref{fig-trans153}.
Also displayed are the series of images of this event.  The top row
represents the precursor/host of the event in its quiescent state, in
the $R$, $I$, and $z'$--bands, from left to right.  Included in each
panel is the magnitude of this precursor/host in its quiescent state,
except for the $I$ band where we have no calibration data.  This
precursor/host is not visible in $B$ or $V$, and appears unresolved in
the $z'$ images but measurably elliptical in both $I$ and $R$.  The
subsequent panels, reading from left to right and top to bottom, show
the emergence of the OT in our time series of images, as well as OT
magnitudes and time elapsed from the detection observation.  All
images are separated by approximately $700s$ except for the final
image, which was obtained more than 2 days after the event.  The event
was also announced via the IAU Circulars
(\mycite{2003IAUC.8094....2B}).

The OT lightcurve shows more complex behavior than the previous
events, and cannot be described as simple power--law flux decay.  The
second temporal peak indicates either a second round of brightening or
spectral evolution of the flare.

As noted above, the precursor/host object appears resolved and
elliptical in the $R$ and $I$ passbands.  The object's position angles
in $R$ and $I$, derived with the adaptive second moment method of
\cite{bj2002}, agree to within $0.09\deg$, indicating an extended
object.  
We have characterized the expected PSF at the location of the host
object by first identifying candidate stars, based on their position
in the size--magnitude plane, in the $R$--band template image.  We
then fit for the spatial variation of the moments of these candidates,
clipping at 3$\sigma$ to reject interloping small galaxies.  A
$\chi^2$ test comparing the shape of the host with the predicted shape
of the PSF at its position yields $\chi^2 = 11.4$ for 3 degrees of
freedom, for a confidence level of $99\%$ that a point source would
not have the measured shape of the precursor object.  We consider the
above as strong evidence for a resolved, extragalactic host for OT
20030305.

\subsection{Discussion}

A primary discriminant between Galactic stellar events, such as OT
20020115, and extragalactic events is the measured shape of the
precursor or host object.  OT 20010326 is shown to be unresolved in
HST imaging, implying Galactic origin, although potentially
originating from a compact extragalactic host.  OT 20030305 has an
apparent resolved host that is inconsistent with the $R$--band PSF
interpolated at its position.  This argues strongly for an
extragalactic origin.

It is suggestive that all of our OTs are detected in the $B$ passband.
We consider various stellar and extragalactic possibilities for the
origin of these events in turn.

\subsubsection{Galactic Flare Stars}

Since we are certain OT 20020115 is from a flaring dwarf star, we
investigate the possibility that one, or even both, of the other OTs
are similar in nature.  We examine the peak of the disk dwarf
luminosity function, which is occupied by dM4 -- dM6 ($M_V = 12 - 16$)
stars \citep{reid-95}.  This region is occupied by the most active UV
Ceti--type flare stars, with event timescales of order minutes and
flare $B - V$ colors of approximately 0.0 to 0.3 \citep{kunkel-75}.
\cite{gurza-flare} further characterizes the ``Type I'' outburst
lightcurves expected of UV Ceti--type stars, with rise times of
seconds to a few minutes, and decay times of minutes to about one
hour, qualitatively similar to our observed events.

\begin{figure*}[t]
  \plotone{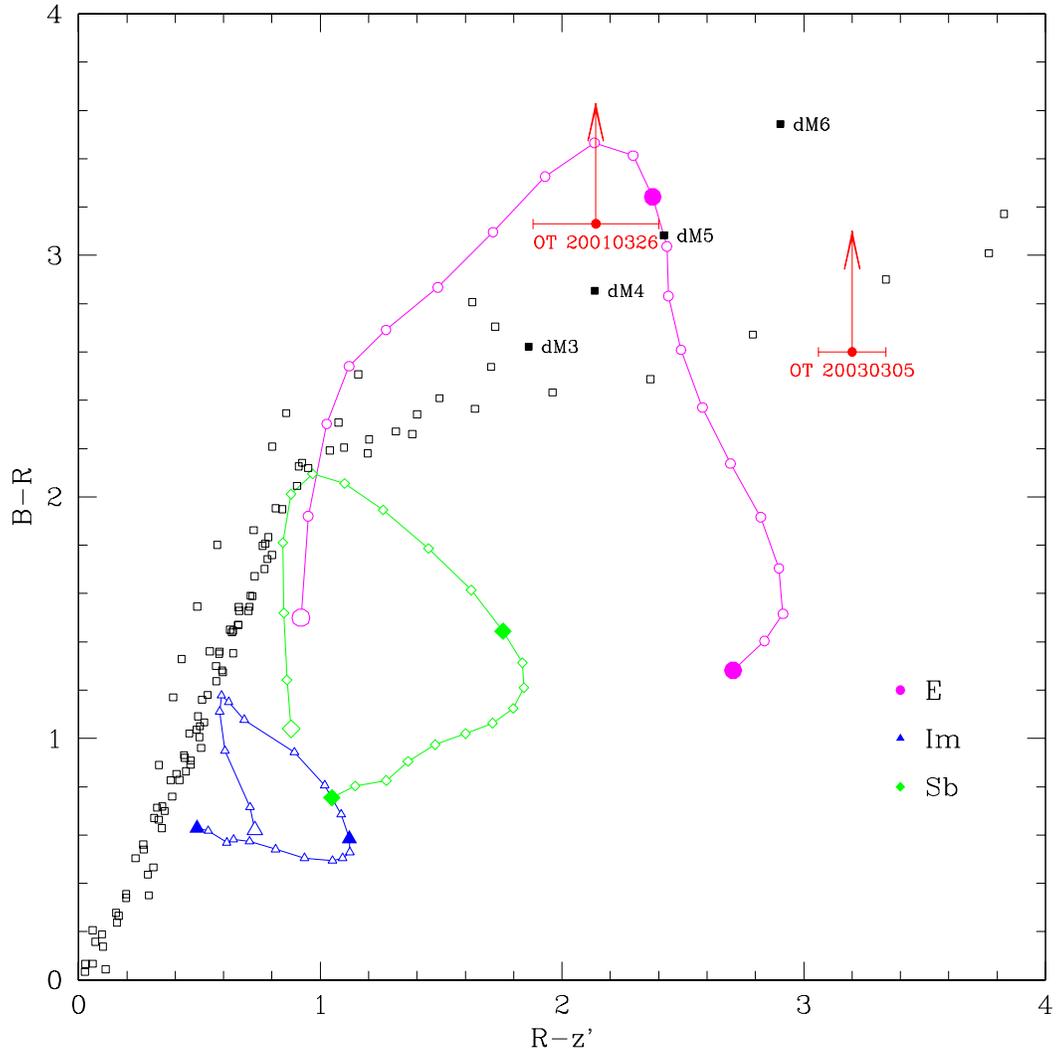}

  \caption{$B-R$ vs. $R-z'$ diagram for the precursor/host objects of
  OT 20010326 and OT 20030305.  The open squares represent the
  \cite{pickles} stellar library of energy distributions, with solid
  squares indicating the approximate positions of dM3--6 dwarf stars.
  The tracks indicate K--corrected colors for elliptical (circle),
  irregular (triangle), and Sb (diamond) galaxies with increasing
  redshift (steps of $z=1$ are indicated by filled symbols, starting
  with the large open symbol at $z=0$) based on galaxy spectral
  templates from \cite{coleman}.  These are shown to indicate regions
  where contamination of stellar precursor identifications by compact
  high redshift galaxies is expected.}

\label{fig-cmds}
\end{figure*}

From Figure~2 of \cite{pat-dwarf}, and Figure~\ref{fig-cmds} here, the
precursor $R - z'$ color of OT 20010326 implies a spectral type of dM4
were it a flare star.  With $B - R > 3.1$, this object lies $1 \sigma$
from the color--defined locus of disk main sequence stars.  However,
using a $V$--band dM4 distance modulus of between 11.5 and 12.5
magnitudes, this object would be approximately 2.0--3.2 kpc distant,
and 1.5--2.4 kpc above the Galactic plane.  This instead suggests
association with the sdM halo subdwarf population, consistent with our
proper motion limits.  However, these stars are not known to flare.
The situation is similar for OT 20030305, which is far too faint at $V
> 27.1$ mag to belong to the disk population.  Assuming an $R - z'$
derived spectral type of approximately dM6, its distance modulus
yields $d \ga 1.7$ kpc.

For a typical DLS field at b(II) = $45\deg$, we expect to find
approximately 100 -- 150 dM4 -- dM6 disk stars per subfield out to
their Galactic scale height of 350 pcs, which is verified through
direct examination of several subfields.  Since flare events do not
follow Poisson statistics, and show large scatter in total energy,
peak light and duration, it is difficult to calculate the exact number
expected in each sequence of B exposures.  In addition, selection
effects in most previous studies have biased discovery towards stars
of greatest flare visibility.

Until recently, one of the only ways to estimate the expected rate of
flare star activity was to extrapolate Equation~5 from
\cite{kunkel-73}.  This relation was derived using data for a selected
few of the most active flare stars in the immediate solar
neighborhood.  Scaling Kunkel's relation to our $B$ passband, and
assuming
that $5\%$ of all dM4--6 stars exhibit strong flaring activity
(i.e. $\Delta B$ of 1 mag or greater), we may place an upper limit of
approximately 0.08 flare events expected per subfield in each 600 sec
exposure.  Given the extreme selection bias in Kunkel's sample, it is
not unexpected to discover that this estimate ($N < 34$) is much
different than what we find ($1 \leq N \leq 3$).  We also examine the
stellar flare survey data of \cite{fresneau-2001}.  Utilizing
astrographic plates covering $520$ deg$^2$ at low Galactic latitude to
apparent $B$ magnitude of 10--14, they find $8\%$ of their stars show
flare events of $> 0.4$ mag over 20--30 minutes.  If half of these
stars are M dwarfs, of which approximately $1\%$ are types dM4--6
based on the volume of the sample limited by the luminosity of those
spectral types, we expect an upper limit of 0.01 flare of $> 0.4$ mag
per 100 stars in each 600 sec $B$--band exposure.
Having analyzed 425 accumulated $B$--band images for this publication,
we expect an upper limit of 6.4 flaring events $> 0.4$ magnitudes.
The distribution function of the amplitudes of flaring events in these
stars is not well known, but in general the brighter flares make up a
smaller fraction of the total number of events \citep{gurza-flare}.
Thus the discovery of only one certain flare event of $\Delta B > 1$
mag is consistent with this upper limit.  Given the lack of unbiased
flare star samples down to our limiting magnitudes, a direct
comparison with other surveys requires the analysis of more modern
variability datasets (e.g FSVS: \mycite{fsvs-03}).

\subsubsection{Active Galactic Nuclei (AGN)}

\cite{sloan-ot1} have reported a highly luminous OT event from the
Sloan Digital Sky Survey, with an underlying host galaxy at $z =
0.385$.  \cite{sloan-ot2} have determined, using in part archival
optical and radio data, that the event is likely due to a radio--loud
AGN.  Heeding the suggestions of \cite{sloan-ot2}, we search archival
data sources for observations of the fields of OT 20010326 and OT
20030305.  We include in these searches the FIRST 1.4 GHz radio
catalog \citep{first-97}, 1.4 GHz NVSS radio catalog \citep{nvss}, and
4.8 GHz PMN catalogs \citep{pmn}.

The only survey that covers the field of OT 20010326 is the PMN
tropical survey, which detects no sources closer than $1.7\arcmin$.
In combination with our VLA detection limits 4 days after the event,
this argues against a radio variable source.  Null detections are also
reported for OT 20030305 in the FIRST, NVSS, and PMN equatorial
catalogs, with an explicit limit from FIRST of 0.94 mJy/beam.
Overall, these null detections seem to reject radio--loud AGN
as sources for this variability.

We have also searched the archival X--ray and gamma--ray catalogs
through HEASARC, and while OT 20010326 returns several matches
corresponding to components of Abell 1836, no matches are found within
$2 \arcmin$.  A search around OT 20030305 finds no matches within $10
\arcmin$.  This weighs against a variety of accretion scenarios,
including AGN and other Galactic X--ray sources.

\subsubsection{Gamma Ray Bursts (GRBs)}

There is an expected population of optical transients associated with
the observed rate of gamma ray bursts.  However, the connection
between the two event rates is very uncertain, given the variety of
means to draw optical emission from hydrodynamic evolution of a GRB
event.  These means include ``orphan'' GRBs, where a highly collimated
GRB points at large enough angle away from the observer that gamma ray
emission is avoided but optical afterglow radiation is not
\citep{rhoads-1997}; and ``dirty'' or ``failed'' fireballs whose
ejecta comprise a significant amount of baryons and/or have a small
Lorentz factor ($\ll$ 100, e.g. \mycite{dermer-grb},
\mycite{hdl-2002}).  Distinguishing between these possibilities
requires fine constraints on the time decay slope $\alpha$, radio
observations, and/or multiwavelength observations at the time of the
so--called GRB ``jet break'' \citep{rhoads-2003}.
Given the our poor temporal resolution, and lack of simultaneous
multi--wavelength coverage, we are not uniquely sensitive to the
micro--physics expected to drive the early evolution of GRB
lightcurves, and whose resolution might distinguish between the above
possibilities.  Within the resolution and timescale of our
observations, all GRB variants effectively yield the same
characteristic isolated fading object.

The thousand second timescale of the observed events is of similar
duration to the early lightcurve break observed in GRB 021211
\citep{li-021211}.  After this break, the power law flux decay index
$\alpha$ was seen to decrease from 1.8 to 0.8 -- before this time, the
transient dimmed by more than 2.5 magnitudes.  If our already faint,
rapidly declining optical transients represent a similar early phase
of gamma--ray dark GRB lightcurves, their subsequent evolution would
be below our detection limit.

We also note that the spectral energy distribution (SED) of the GRB
afterglow component is modeled as a synchrotron emission spectrum
(e.g. \mycite{sari-spec}, \mycite{sari-spec2}).  The expected spectral
shape scales like $F_\nu \propto \nu^{-\beta}$.  Thus it is surprising
that our distribution of $B$:$V$:$R$--detected transients is 2:1:0
(where OT 20030305 contributes to both the $B$ and $V$ event rate) if
these events are drawn from a population of events characterized by a
synchrotron SED.

Recent submillimeter and radio observations of localized GRB host
galaxies show they are bluer than selected galaxies at similar
redshifts, indicative of star formation or relatively low dust content
(e.g. \mycite{berger-host}, \mycite{floch-grbhost}).  Host galaxies
for localized X--ray Flashes (XRFs) also appear to exhibit the same
characteristics \citep{bloom-2003}.  However, our putative hosts are
considerably redder than these samples of GRB or XRF--selected host
galaxies, and are unlikely to be drawn from the same population.  The
$R - z'$ colors of our OT hosts are, however, loosely consistent with
the observed population of Extremely Red Objects (EROs), which
comprise old or dusty star--forming galaxies.  If EROs represent
highly obscured starburst activity at moderate redshift ($z \sim 1$),
they contribute a significant fraction of the overall star formation
at that epoch \citep{smail-ero}.  Assuming our population of OTs
traces star formation, we expect some association with EROs.  The
absence of extremely red GRB host galaxies suggests detection of GRBs
is biased against dusty galaxies.  If our OTs arise from dust
enshrouded galaxies, this bias would seem lesser for optically
detected events.

\section{Event Rate}

We map our detection of these objects into a quantitative statement
describing the rate of short timescale astrophysical variability.  We
derive general constraints without bias towards a particular OT model.

\subsection{Efficiency}

To account for inefficiencies in our pipeline, we model our efficiency
$\eff$ at recovering transients throughout a range in brightness.  We
calibrate the efficiency via Monte Carlo runs in which we add point
source transients.  We define an efficiency run as the generation of
32 random amplitude point sources, with the integrated flux chosen as
a uniform deviate in the exponent of $f = 10^{1.5 \leq x \leq 6}$,
which are placed within a dither sequence of survey data.  Positions
are randomly selected within the limits of the primary exposure, and
are placed at the same astronomical position within each of the
subsequent 4 dithers.  We do not bias our results by assuming a
temporal shape for the variability.  Instead, we generate an overall
efficiency for recovering a given amount of input flux, averaged over
the spatial coverage of our dither sequence.

We include in our analysis $B$, $V$, and $R$--band survey data
($z'$--band data are infrequently reduced in real--time due to
fringing complications).  We account for variations in observing
conditions by choosing 19 night--filter--subfield combinations of
survey data to submit for efficiency analysis.  In total, we initiated
$8028$ efficiency runs yielding $256896$ total efficiency points,
approximately 85000 per passband.  The magnitudes of these transients
are calibrated to the \cite{landolt-92} system.  Image zero--point
offsets were calculated from directly calibrated images of each
subfield.  One subfield was calibrated using publicly available NOAO
Deep Wide--Field Survey data \citep{noao-deepwide}.

\begin{figure*}[t]
  \plotone{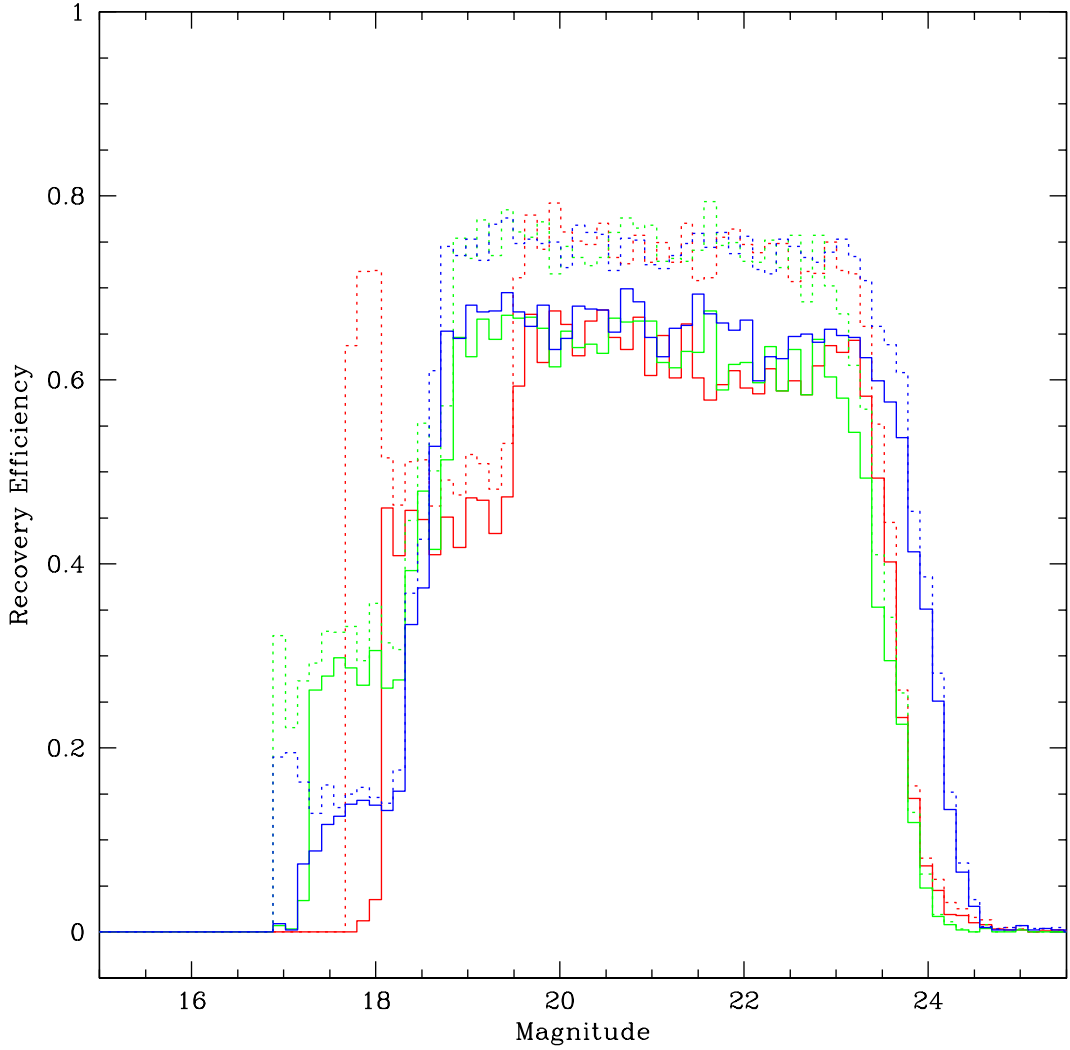}

  \caption{Transient point--source detection efficiency as a function
    of magnitude in the difference image ($\mathcal{M}$), for the $B$,
    $V$, and $R$ passbands.  The {\it dotted} histograms are our raw
    point source detection efficiencies, while the {\it solid}
    histograms represent our efficiencies after cuts.  The bright end
    cutoff represents the level at which the objects become saturated,
    with additional efficiency coming from the recovery of the wings
    of the saturated object.  In our efficiency analysis, we explicity
    set our efficiency for objects brighter than saturation
    ($\mathcal{M}_B \leq 18.6, \mathcal{M}_V \leq 18.8$ and
    $\mathcal{M}_R \leq 19.5$) equal to zero.  The dim cutoff
    represents the 2.5 $\sigma$ detection limit of single images in
    our transient search.  }

\label{fig-eff}
\end{figure*}

After placing additional flux in each image, we pass them through our
difference imaging pipeline.  The number of input objects recovered is
tallied, searching both the unfiltered and filtered object catalogs
generated at the {\tt FI} stage in our pipeline.  This yields an
unfiltered detection efficiency ({\it dotted} histogram in
Figure~\ref{fig-eff}) as well as the efficiency of passing our cuts
({\it solid} histogram in Figure~\ref{fig-eff}).  
The difference between the two histograms demonstrates the rejection
of true positives in our filtering process, yielding an overall
decrease in efficiency.  We use the {\it solid} histogram in the
subsequent event rate analysis.
%
%
We note that approximately $3\%$ of our sky coverage lands in gaps
between MOSAIC chips, and the dithering procedure shifts $10\%$ and
$20\%$ of the template images' common sky area off of the subsequent
dithers.  Thus our maximum possible efficiency is $\sim 82\%$.  Our
overall OT detection efficiency of $\eff \sim 63\%$ indicates our
filtering efficiency is $\sim 77\%$.  We expect substantial
improvements in the context of a future automated classification
environment.

Figure~\ref{fig-eff} shows our efficiency is fairly constant between
the saturation limit on the bright end, and our detection limits on
the faint end.  We thus expect little bias against faint transient
detection in this given range.  On the bright end, there appears to be
non--zero $\eff$ for objects brighter than saturation.  These are
systematic artifacts that result from the detection of unsaturated and
unmasked wings of saturated efficiency objects, in close enough
proximity to the input object itself to warrant a positional match.
In the following analysis, we set $\eff \equiv 0$ brighter than the
average saturation limits of $\mathcal{M}_B = 18.6, \mathcal{M}_V =
18.8$ and $\mathcal{M}_R = 19.5$.

\subsection{General Constraints on Variability}

To average over unknown inefficiencies in the human element of our
transient pipeline, we require that a transient be confirmed in a
subsequent image, which we consider a human observer $100\%$ efficient
at classifying as real.  Thus our limits on short timescale
variability correspond to twice our typical exposure time, plus $100s$
of readout between images.  Our efficiencies in Figure~\ref{fig-eff}
are relatively constant between the bright and faint limits, which we
can approximate by a constant efficiency $\left< \eff \right>$ without
loss of specificity.  However, due to the differing $\left< \eff
\right>$ per passband, differences in typical exposure times, and in
number of events detected, we also quote event rates for each passband
individually.  The observed rate for each passband, averaged over all
combinations of detectable variability amplitudes, is
\begin{eqnarray}
\rate = {N \over \left<\eff\right>^2 ~ \exposure}~{\rm events~deg^{-2}~day^{-1}}
\end{eqnarray}
where $N$ is the number of observed events, $\exposure$ the
appropriate exposure, and $\left<\eff\right>$ is the averaged
efficiency, squared due to the requirement of two detections.  We note
that OT 20030305 contributes to both the $V$ and $B$ event rate, as it
was discovered in both sets of images.  

For the $B$ passband with 3 detected transients, our overall rate for
variability on $1300s$ timescales, between $18.6 < \mathcal{M}_B <
23.8$, is $\rate = 6.5$ events deg$^{-2}~{\rm day}^{-1}$.  We
recognize at least one of these events as Galactic in origin.  Having
detected no more than two cosmological events, Poisson statistics
exclude at the $95\%$ level any OT model that predicts a mean number
of detectable cosmological events $N_B > 6.3$, implying $\rate_B < 14$
events deg$^{-2}~{\rm day}^{-1}$.  Table~\ref{tab-rate} lists, for a
given number of considered events, experimental rates and limits on
$\rate$.  We have detected zero short timescale events in the $R$
passband, constraining overall $1900s$ astronomical variability
between $19.5 < \mathcal{M}_R < 23.4$ to $\rate_{95\%} < 5.2$ events
deg$^{-2}~{\rm day}^{-1}$.

In our total summed exposure of $3.7$ deg$^2$--days from all
passbands, assuming $\left< \eff \right> = 0.63$ and 4 events (3
unique events), we find an overall rate of short timescale
astronomical variability of $\rate = 2.7~(2.0)~{\rm events}~{\rm
deg}^{-2}~{\rm day}^{-1}$, with $\rate_{95\%} < 6.3~(5.3)$.  These are
the first general constraints on short timescale variability at such
depths.

\section{Conclusions}

We have reported on the structure of and first results from our
wide--field image subtraction pipeline.  An important characterization
of our transient survey is the exposure $\exposure$ at a given
timescale and to a given depth.  The DLS transient search is primarily
sensitive to $\sim 1000s$ variability from $19^{th}$ to deeper than
$23^{rd}$ magnitudes in $B$, $V$, and $R$.  Within this envelope of
sensitivity, we have detected three short timescale optical transient
events.

OT 20010326 occurred in the field of galaxy cluster Abell 1836.  One
archival HST image of the cluster includes this region, and the
transient precursor is present and unresolved.  This indicates a
compact precursor, stellar in nature if it resides in our Galaxy.  The
colors of this object are consistent at the $1 \sigma$ level with
those expected of Galactic dwarf stars (Figure~\ref{fig-cmds}), whose
flaring activity presents a known background.  However, the object
would be too far out of the Galactic plane to belong to the disk
population of dwarf stars, and halo subdwarfs are not known to flare.
It is also possible the precursor resides in, or behind, the Abell
cluster, but overall its nature remains uncertain.  OT 20020115 is
identified as a Galactic M dwarf of spectral type dM4, exhibiting
classical flare star activity.  Finally, the host for OT 20030305
appears consistently elliptical in the $R$ and $I$ passbands, and
inconsistent with the $R$--band stellar PSF, which we consider a
strong argument for a resolved host, extragalactic in nature.  We find
OT 20030305 the strongest candidate so far for optically detected,
short timescale cosmological variability.

The precursor or host objects for OTs 20010326 and 20030305 are
definitively redder than galaxies that are known to host GRBs and
XRFs.  If our events are cosmological in nature, this suggests that
optical and high energy events arise from different mechanisms, or,
given the dearth of GRBs from dusty star--forming galaxies, a stronger
bias against GRBs and XRFs from a dusty environment.  If our
lightcurves evolve analogous to the prompt stage of GRBs emission, the
rapid decline coupled with their intrinsic faintness would make them
difficult to monitor beyond several hours.  Finally, we emphasize the
diversity of variable objects in the stellar menagerie, and thus it is
not straightforward to rule out Galactic stars as the sources for our
OTs.  Spectroscopic information will ultimately help to clarify their
nature, and such followup observations are planned.

Our search has also yielded SN--like events that appear to have no
host galaxy to significant ($> 27$) limiting magnitude -- OT 20020112
is one such example.  In addition, our catalog of SN candidates,
classified as supernovae primarily due to their proximity to a host
galaxy, might also contain sources with unusual temporal evolution.

Overall, the DLS transient search is well suited to explore the
parameter space of OTs with small energy budgets, a population that
could plausibly have escaped detection by gamma--ray and X--ray
satellite missions.  This raises the possibility that the phenomena
detected represent a new class of astronomical variability.
Coordinated photometric followup of future optical transients is
absolutely necessary to reveal fully the diversity of variability at
faint optical magnitudes.  A primary goal of future variability
surveys must be to enable the photometric and spectroscopic followup
of detected events through timely release of information and ease of
access to available data.

The wealth of information that can be gleaned from real--time synoptic
transient science, only a subset of which is covered in this paper,
strengthens the science case for the expansion of deep, wide
astronomical surveys into the short timescale regime.  Expected future
surveys like the LSST and PAN--STARRS \citep{pan-starrs} will survey
thousands of square degrees per night, and will benefit greatly from
modern development in this field.

\section*{Acknowledgments}

We thank the many observers who have assisted in reviewing transients
during the course of our survey, including M. Lopez-Caniego Alcarria,
P. Baca, H. Khiabanian, J. Kubo, D. Loomba, C. Navarro, and R. Wilcox.
We thank G. Bernstein for orbit calculations, R. Sari and S. Hawley
for useful discussions, and G. Bravo, M. Hamui, D. Kirkman, C. Smith,
and C. Stubbs for assistance.  We are very grateful for the skilled
support of the staff at CTIO and KPNO.  The Deep Lens Survey is
supported by NSF grants AST-0307714 and AST-0134753.  AC acknowledges
the support of CONICYT, through grant FONDECYT 1000524.  This work
made use of images and/or data products provided by the NOAO Deep
Wide-Field Survey, which is supported by the National Optical
Astronomy Observatory (NOAO). NOAO is operated by AURA, Inc., under a
cooperative agreement with the National Science Foundation.  This
research has made use of NASA's Astrophysics Data System Bibliographic
Services.  This work made use of images obtained with the Magellan I
(Baade) Telescope, operated by the Observatories of the Carnegie
Institution of Washington.  Includes observations made with the
NASA/ESA Hubble Space Telescope, obtained from the data archive at the
Space Telescope Science Institute. STScI is operated by the
Association of Universities for Research in Astronomy, Inc. under NASA
contract NAS 5-26555.



\begin{thebibliography}{84}
\expandafter\ifx\csname natexlab\endcsname\relax\def\natexlab#1{#1}\fi

\bibitem[{{Afonso} {et~al.}(2003){Afonso}, {Albert}, {Andersen}, {Ansari},
  {Aubourg}, {Bareyre}, {Beaulieu}, {Blanc}, {Charlot}, {Couchot}, {Coutures},
  {Ferlet}, {Fouqu{\' e}}, {Glicenstein}, {Goldman}, {Gould}, {Graff}, {Gros},
  {Haissinski}, {Hamadache}, {de Kat}, {Lasserre}, {Le Guillou}, {Lesquoy},
  {Loup}, {Magneville}, {Marquette}, {Maurice}, {Maury}, {Milsztajn}, {Moniez},
  {Palanque-Delabrouille}, {Perdereau}, {Pr{\' e}vot}, {Rahal}, {Rich},
  {Spiro}, {Tisserand}, {Vidal-Madjar}, {Vigroux}, \&
  {Zylberajch}}]{eros-smc03}
{Afonso}, C., {et~al.} 2003, \aap, 400, 951

\bibitem[{{Akerlof} {et~al.}(1993){Akerlof}, {Fatuzzo}, {Lee},
  {Bionta}, {Ledebuhr}, {Park}, {Barthelmy}, {Cline}, \&
  {Gehrels}}]{grocse} 
{Akerlof}, C., {et~al.} 1993, in BATSE Gamma Ray Burst Workshop, 21--23, 2

\bibitem[{{Akerlof} {et~al.}(2003){Akerlof}, {Kehoe}, {McKay}, {Rykoff},
  {Smith}, {Casperson}, {McGowan}, {Vestrand}, {Wozniak}, {Wren}, {Ashley},
  {Phillips}, {Marshall}, {Epps}, \& {Schier}}]{rotse-3}
{Akerlof}, C.~W., {et~al.} 2003, \pasp, 115, 132

\bibitem[{{Alard}(2000)}]{alard-isis00}
{Alard}, C. 2000, \aaps, 144, 363

\bibitem[{{Alcock} {et~al.}(2000{\natexlab{b}}){Alcock}, {Allsman}, {Alves},
  {Axelrod}, {Becker}, {Bennett}, {Cook}, {Dalal}, {Drake}, {Freeman}, {Geha},
  {Griest}, {Lehner}, {Marshall}, {Minniti}, {Nelson}, {Peterson}, {Popowski},
  {Pratt}, {Quinn}, {Stubbs}, {Sutherland}, {Tomaney}, {Vandehei}, \&
  {Welch}}]{macho-lmc2000}
{Alcock}, C., {et~al.} 2000{\natexlab{b}}, \apj, 542, 281

\bibitem[{{Aldering} {et~al.}(2002){Aldering}, {Adam}, {Antilogus}, {Astier},
  {Bacon}, {Bongard}, {Bonnaud}, {Copin}, {Hardin}, {Henault}, {Howell},
  {Lemonnier}, {Levy}, {Loken}, {Nugent}, {Pain}, {Pecontal}, {Pecontal},
  {Perlmutter}, {Quimby}, {Schahmaneche}, {Smadja}, \& {Wood-Vasey}}]{snfac}
{Aldering}, G., {Adam}, G., {Antilogus}, P., {Astier}, P., {Bacon}, R.,
  {Bongard}, S., {Bonnaud}, C., {Copin}, Y., {Hardin}, D., {Henault}, F.,
  {Howell}, D.~A., {Lemonnier}, J., {Levy}, J., {Loken}, S.~C., {Nugent},
  P.~E., {Pain}, R., {Pecontal}, A., {Pecontal}, E., {Perlmutter}, S.,
  {Quimby}, R.~M., {Schahmaneche}, K., {Smadja}, G., \& {Wood-Vasey}, W.~M.
  2002, in Survey and Other Telescope Technologies and Discoveries. Edited by
  Tyson, J. Anthony; Wolff, Sidney. Proceedings of the SPIE, Volume 4836, pp.
  61-72 (2002)., 61--72

\bibitem[{{Alves} {et~al.}(2003){Alves}, {Bergier}, {Crotts}, {Cseresnjes}, \&
  {Gersch}}]{mega-03}
{Alves}, D.~R., {et~al.} 2003, in IAU Symposium 220

\bibitem[{{Barris} {et~al.}(2003){Barris}, {Smith}, {Aguilera}, {Krisciunas},
  {Suntzeff}, {Becker}, {Covarrubias}, {Miceli}, {Miknaitis}, {Rest}, {Stubbs},
  {Garnavich}, {Holland}, {Schmidt}, {Filippenko}, {Jha}, {Li}, {Challis},
  {Kirshner}, {Matheson}, {Tonry}, {Riess}, {Leibundgut}, {Sollerman},
  {Spyromilio}, {Clocchiatti}, \& {Pompea}}]{barris-ifa}
{Barris}, B., {et~al.} 2003, Submitted

\bibitem[{{Becker}(2003{\natexlab{a}})}]{2003IAUC.8093....2B}
{Becker}, A. 2003{\natexlab{a}}, \iaucirc, 8093, 2

\bibitem[{{Becker}(2003{\natexlab{b}})}]{2003IAUC.8094....2B}
---. 2003{\natexlab{b}}, \iaucirc, 8094, 2

\bibitem[{{Becker} {et~al.}(2002{\natexlab{a}}){Becker}, {Baca},
  {Dell'Antonio}, \& {Norman}}]{2002IAUC.7833....1B}
{Becker}, A., {et~al.} 2002{\natexlab{a}}, \iaucirc, 7833, 1

\bibitem[{{Becker} {et~al.}(2003){Becker}, {Loomba}, \&
  {Wilcox}}]{2003IAUC.8093....1B}
{Becker}, A., {Loomba}, D., \& {Wilcox}, R. 2003, \iaucirc, 8093, 1

\bibitem[{{Becker} {et~al.}(2002{\natexlab{b}}){Becker}, {Lopez-Caniego},
  {Norman}, {Wittman}, {Halpern}, \& {Jackson}}]{2002IAUC.7803....2B}
{Becker}, A., {et~al.} 2002{\natexlab{b}}, \iaucirc, 7803, 2

\bibitem[{{Becker} {et~al.}(2002{\natexlab{c}}){Becker}, {Tyson}, {Wittman},
  {Lopez-Caniego}, {Norman}, {Clocchiatti}, {Dell'Antonio}, \&
  {Squires}}]{2002IAUC.7804....1B}
{Becker}, A., {et~al.} 2002{\natexlab{c}}, \iaucirc, 7804, 1

\bibitem[{{Becker} {et~al.}(2002{\natexlab{d}}){Becker}, {Wittman}, {Tyson}, \&
  {Clocchiatti}}]{2002GCN..1217....1B}
{Becker}, A.~C., {et~al.} 2002{\natexlab{d}}, GRB Circular Network, 1217, 1

\bibitem[{{Berger} {et~al.}(2003){Berger}, {Cowie}, {Kulkarni}, {Frail},
  {Aussel}, \& {Barger}}]{berger-host}
{Berger}, E., {et~al.} 2003, \apj, 588, 99

\bibitem[{{Bernstein} \& {Khushalani}(2000)}]{bern-kbofit}
{Bernstein}, G. \& {Khushalani}, B. 2000, \aj, 120, 3323

\bibitem[{{Bernstein} \& {Jarvis}(2002)}]{bj2002}
{Bernstein}, G.~M. \& {Jarvis}, M. 2002, \aj, 123, 583

\bibitem[{{Bertin} \& {Arnouts}(1996)}]{bertin96}
{Bertin}, E. \& {Arnouts}, S. 1996, \aaps, 117, 393

\bibitem[{{Bloom} {et~al.}(2003){Bloom}, {Fox}, {van Dokkum}, {Kulkarni},
  {Berger}, {Djorgovski}, \& {Frail}}]{bloom-2003}
{Bloom}, J.~S., {et~al.} 2003, \apj, 599, 957

\bibitem[{{Boeshaar} {et~al.}(2003){Boeshaar}, {Margoniner}, \& {The Deep Lens
  Survey Team}}]{pat-dwarf}
{Boeshaar}, P.~C., {Margoniner}, V., \& {The Deep Lens Survey Team}. 2003, in
  IAU Symposium 211, E. Martin, ed., Astron. Soc. Pacific. 203

\bibitem[{{Clocchiatti} {et~al.}(2002){Clocchiatti}, {Boeshaar}, {Becker}, \&
  {Wittman}}]{2002GCN..1218....1C}
{Clocchiatti}, A.,  {et~al.} 2002, GRB Circular Network, 1218, 1

\bibitem[{{Coleman} {et~al.}(1980){Coleman}, {Wu}, \& {Weedman}}]{coleman}
{Coleman}, G.~D., {Wu}, C.-C., \& {Weedman}, D.~W. 1980, \apjs, 43, 393

\bibitem[{{Condon} {et~al.}(1998){Condon}, {Cotton}, {Greisen}, {Yin},
  {Perley}, {Taylor}, \& {Broderick}}]{nvss}
{Condon}, J.~J., {et~al.} 1998, \aj, 115, 1693

\bibitem[{{Dermer} {et~al.}(1999){Dermer}, {Chiang}, \& {B{\"
  o}ttcher}}]{dermer-grb}
{Dermer}, C.~D., {Chiang}, J., \& {B{\" o}ttcher}, M. 1999, \apj, 513, 656

\bibitem[{{Filippenko} {et~al.}(2001){Filippenko}, {Li}, {Treffers}, \& {et
  al.}}]{loss}
{Filippenko}, A.~V., {et~al.} 2001, in ASP Conf. Ser. 246: IAU Colloq. 183: Small Telescope Astronomy on Global Scales, 121--+

\bibitem[{{Fresneau} {et~al.}(2001){Fresneau}, {Argyle}, {Marino}, \&
  {Messina}}]{fresneau-2001}
{Fresneau}, A., {et~al.} 2001, \aj, 121, 517

\bibitem[{{Gal-Yam} {et~al.}(2002){Gal-Yam}, {Ofek}, {Filippenko}, {Chornock},
  \& {Li}}]{sloan-ot2}
{Gal-Yam}, A., {et~al.} 2002, \pasp, 114, 587

\bibitem[{{Granot} {et~al.}(2000){Granot}, {Piran}, \& {Sari}}]{sari-spec2}
{Granot}, J., {Piran}, T., \& {Sari}, R. 2000, \apjl, 534, L163

\bibitem[{{Griffith} \& {Wright}(1993)}]{pmn}
{Griffith}, M.~R. \& {Wright}, A.~E. 1993, \aj, 105, 1666

\bibitem[{{Groot} {et~al.}(2003){Groot}, {Vreeswijk}, {Huber}, {Everett},
  {Howell}, {Nelemans}, {van Paradijs}, {van den Heuvel}, {Augusteijn},
  {Kuulkers}, {Rutten}, \& {Storm}}]{fsvs-03}
{Groot}, P.~J., {et~al.} 2003, \mnras, 339, 427

\bibitem[{{Gurzadian}(1980)}]{gurza-flare}
{Gurzadian}, G.~A. 1980, Oxford Pergamon Press International Series on Natural
  Philosophy, 101

\bibitem[{{Howell} {et~al.}(1996){Howell}, {Koehn}, {Bowell}, \&
  {Hoffman}}]{loneos}
{Howell}, S.~B., {et~al.} 1996, \aj, 112, 1302

\bibitem[{{Huang} {et~al.}(2002){Huang}, {Dai}, \& {Lu}}]{hdl-2002}
{Huang}, Y.~F., {Dai}, Z.~G., \& {Lu}, T. 2002, \mnras, 332, 735

\bibitem[{{Ivezi{\' c}} {et~al.}(2003){Ivezi{\' c}}, {Lupton}, {Anderson},
  {Eyer}, {Gunn}, {Juri{\' c}}, {Knapp}, {Miknaitis}, {Gunn}, {Rockosi},
  {Schlegel}, {Strauss}, {Stubbs}, \& {Vanden Berk}}]{ivesic-sdsstemporal}
{Ivezi{\' c}}, {\v Z}., {et~al.} 2003, Memorie della Societa Astronomica Italiana, 74, 978

\bibitem[{{Ivezi{\' c}} {et~al.}(2001){Ivezi{\' c}}, {Tabachnik}, {Rafikov},
  {Lupton}, {Quinn}, {Hammergren}, {Eyer}, {Chu}, {Armstrong}, {Fan},
  {Finlator}, {Geballe}, {Gunn}, {Hennessy}, {Knapp}, {Leggett}, {Munn},
  {Pier}, {Rockosi}, {Schneider}, {Strauss}, {Yanny}, {Brinkmann}, {Csabai},
  {Hindsley}, {Kent}, {Lamb}, {Margon}, {McKay}, {Smith}, {Waddel}, {York}, \&
  {the SDSS Collaboration}}]{ivezic-ast}
{Ivezi{\' c}}, {\v Z}., {et~al.} 2001, \aj, 122, 2749

\bibitem[{{Jannuzi} \& {Dey}(1999)}]{noao-deepwide}
{Jannuzi}, B.~T. \& {Dey}, A. 1999, in ASP Conf. Ser. 191: Photometric
  Redshifts and the Detection of High Redshift Galaxies, 111--+

\bibitem[{{Kaiser} {et~al.}(2002){Kaiser}, {Aussel}, {Burke}, {Boesgaard},
  {Chambers}, {Chun}, {Heasley}, {Hodapp}, {Hunt}, {Jedicke}, {Jewitt},
  {Kudritzki}, {Luppino}, {Maberry}, {Magnier}, {Monet}, {Onaka}, {Pickles},
  {Rhoads}, {Simon}, {Szalay}, {Szapudi}, {Tholen}, {Tonry}, {Waterson}, \&
  {Wick}}]{pan-starrs}
{Kaiser}, N., {et~al.} 2002, in Survey and Other Telescope
  Technologies and Discoveries. Edited by Tyson, J. Anthony; Wolff, Sidney.
  Proceedings of the SPIE, Volume 4836, pp. 154-164 (2002)., 154--164

\bibitem[{{Kirkman} {et~al.}(2000){Kirkman}, {Dell'Antonio}, {Schommer},
  {Squires}, {Tyson}, {Wittman}, {Becker}, {Stubbs}, \&
  {Bernstein}}]{2000IAUC.7398....2K}
{Kirkman}, D., {et~al.} 2000, \iaucirc, 7398, 2

\bibitem[{{Knop} {et~al.}(2003){Knop}, {Aldering}, {Amanullah}, {Astier},
  {Blanc}, {Burns}, {Conley}, {Deustua}, {Doi}, {Ellis}, {Fabbro}, {Folatelli},
  {Fruchter}, {Garavini}, {Garmond}, {Garton}, {Gibbons}, {Goldhaber},
  {Goobar}, {Groom}, {Hardin}, {Hook}, {Howell}, {Kim}, {Lee}, {Lidman},
  {Mendez}, {Nobili}, {Nugent}, {Pain}, {Panagia}, {Pennypacker}, {Perlmutter},
  {Quimby}, {Raux}, {Regnault}, {Ruiz-Lapuente}, {Sainton}, {Schaefer},
  {Schahmaneche}, {Smith}, {Spadafora}, {Stanishev}, {Sullivan}, {Walton},
  {Wang}, {Wood-Vasey}, \& {Yasuda}}]{scp-03}
{Knop}, R.~A., {et~al.} 2003, \apj, 598, 102

\bibitem[{{Kunkel}(1973)}]{kunkel-73}
{Kunkel}, W.~E. 1973, \apjs, 25, 1

\bibitem[{{Kunkel}(1975)}]{kunkel-75}
{Kunkel}, W.~E. 1975, in IAU Symp. 67: Variable Stars and Stellar Evolution,
  15--46

\bibitem[{{Landolt}(1992)}]{landolt-92}
{Landolt}, A.~U. 1992, \aj, 104, 340

\bibitem[{{Le Floc'h} {et~al.}(2003){Le Floc'h}, {Duc}, {Mirabel}, {Sanders},
  {Bosch}, {Diaz}, {Donzelli}, {Rodrigues}, {Courvoisier}, {Greiner},
  {Mereghetti}, {Melnick}, {Maza}, \& {Minniti}}]{floch-grbhost}
{Le Floc'h}, E., {et~al.} 2003, \aap, 400, 499

\bibitem[{{Li} \& {Paczy{\' n}ski}(1998)}]{li-pac}
{Li}, L. \& {Paczy{\' n}ski}, B. 1998, \apjl, 507, L59

\bibitem[{{Li} {et~al.}(2003){Li}, {Filippenko}, {Chornock}, \&
  {Jha}}]{li-021211}
{Li}, W., {et~al.} 2003, \apjl, 586,  L9

\bibitem[{{Mahabal} {et~al.}(2003){Mahabal}, {Djorgovski}, {Graham},
  {Williams}, {Granett}, {Bogosavljevic}, {Baltay}, {Rabinowitz}, {Bauer},
  {Andrews}, {Morgan}, {Snyder}, {Ellman}, {Brunner}, {Rengstorf}, {Musser},
  {Gebhard}, \& {Mufson}}]{palomar-quest}
{Mahabal}, A., {et~al.} 2003, American Astronomical Society Meeting, 203,

\bibitem[{{Malkan} {et~al.}(1998){Malkan}, {Gorjian}, \& {Tam}}]{malkan-agn}
{Malkan}, M.~A., {Gorjian}, V., \& {Tam}, R. 1998, \apjs, 117, 25

\bibitem[{{Millis} {et~al.}(2002){Millis}, {Buie}, {Wasserman}, {Elliot},
  {Kern}, \& {Wagner}}]{des}
{Millis}, R.~L., {et~al.} 2002, \aj, 123, 2083

\bibitem[{{Monet} {et~al.}(1998){Monet}, {Canzian}, {Dahn}, {Guetter},
  {Harris}, {Henden}, {Levine}, {Luginbuhl}, {Monet}, {Rhodes}, {Riepe},
  {Sell}, {Stone}, {Vrba}, \& {Walker}}]{usno-v2}
{Monet}, D.~B.~A., {et~al.} 1998, VizieR Online Data Catalog, 1252, 0

\bibitem[{{Nakar} {et~al.}(2002){Nakar}, {Piran}, \& {Granot}}]{nakar-2002}
{Nakar}, E., {Piran}, T., \& {Granot}, J. 2002, \apj, 579, 699

\bibitem[{{Nemiroff}(2003)}]{nemiroff-03}
{Nemiroff}, R.~J. 2003, \aj, 125, 2740

\bibitem[{{Paulin-Henriksson} {et~al.}(2003){Paulin-Henriksson}, {Baillon},
  {Bouquet}, {Carr}, {Cr{\' e}z{\' e}}, {Evans}, {Giraud-H{\' e}raud}, {Gould},
  {Hewett}, {Kaplan}, {Kerins}, {Le Du}, {Melchior}, {Smartt}, {Valls-Gabaud},
  \& {The POINT-AGAPE Collaboration}}]{pointagape-03}
{Paulin-Henriksson}, S., {et~al.} 2003, \aap, 405, 15

\bibitem[{Perkins {et~al.}(2000)Perkins, Theiler, Brumby, Harvey, Porter,
  Szymanski, \& Bloch}]{genie}
Perkins, S., {et~al.} 2000, in Proceedings of SPIE, Vol. 4120, 52--62

\bibitem[{{Pickles}(1998)}]{pickles}
{Pickles}, A.~J. 1998, \pasp, 110, 863

\bibitem[{{Pojma{\' n}ski}(2001)}]{asas-3}
{Pojma{\' n}ski}, G. 2001, in ASP Conf. Ser. 246: IAU Colloq. 183: Small
  Telescope Astronomy on Global Scales, 53--+

\bibitem[{{Pravdo} {et~al.}(1999){Pravdo}, {Rabinowitz}, {Helin}, {Lawrence},
  {Bambery}, {Clark}, {Groom}, {Levin}, {Lorre}, {Shaklan}, {Kervin},
  {Africano}, {Sydney}, \& {Soohoo}}]{neat-asteroid}
{Pravdo}, S.~H., {et~al.} 1999, \aj, 117, 1616

\bibitem[{{Reid} {et~al.}(1995){Reid}, {Hawley}, \& {Gizis}}]{reid-95}
{Reid}, I.~N., {Hawley}, S.~L., \& {Gizis}, J.~E. 1995, \aj, 110, 1838

\bibitem[{{Rest} {et~al.}(2004){Rest}, {Miceli}, {Miknaitis}, {Covarrubias},
  {Becker}, {Smith}, {Olsen}, {Cook}, {Nikolaev}, {Keller}, {Proctor}, \&
  {Welch}}]{supermacho-03}
{Rest}, A., {et~al.} 2004, In preparation

\bibitem[{{Rhoads}(1997)}]{rhoads-1997}
{Rhoads}, J.~E. 1997, \apjl, 487, L1+

\bibitem[{{Rhoads}(2003)}]{rhoads-2003}
---. 2003, \apj, 591, 1097

\bibitem[{{Rose} {et~al.}(1995){Rose}, {Akella}, {Binegar}, {Choo},
  {Heller-Boyer}, {Hester}, {Hyde}, {Perrine}, {Rose}, \& {Steuerman}}]{opus}
{Rose}, J., {et~al.} 1995, in ASP Conf. Ser. 77: Astronomical Data Analysis Software and Systems
  IV, 429--+

\bibitem[{{Sari} {et~al.}(1998){Sari}, {Piran}, \& {Narayan}}]{sari-spec}
{Sari}, R., {Piran}, T., \& {Narayan}, R. 1998, \apjl, 497, L17+

\bibitem[{{Schlegel} {et~al.}(1998){Schlegel}, {Finkbeiner}, \&
  {Davis}}]{schlegel}
{Schlegel}, D.~J., {Finkbeiner}, D.~P., \& {Davis}, M. 1998, \apj, 500, 525+

\bibitem[{{Smail} {et~al.}(2002){Smail}, {Owen}, {Morrison}, {Keel}, {Ivison},
  \& {Ledlow}}]{smail-ero}
{Smail}, I., {et~al.} 2002, \apj, 581, 844

\bibitem[{{Smith} {et~al.}(2002{\natexlab{a}}){Smith}, {Tucker}, {Kent},
  {Richmond}, {Fukugita}, {Ichikawa}, {Ichikawa}, {Jorgensen}, {Uomoto},
  {Gunn}, {Hamabe}, {Watanabe}, {Tolea}, {Henden}, {Annis}, {Pier}, {McKay},
  {Brinkmann}, {Chen}, {Holtzman}, {Shimasaku}, \& {York}}]{sloan-cal}
{Smith}, J.~A., {et~al.} 2002{\natexlab{a}}, \aj, 123, 2121

\bibitem[{{Smith} {et~al.}(2002{\natexlab{b}}){Smith}, {Aguilera},
  {Krisciunas}, {Suntzeff}, {Becker}, {Covarrubias}, {Miceli}, {Miknaitis},
  {Rest}, {Stubbs}, {Garnavich}, {Holland}, {Schmidt}, {Filippenko}, {Jha},
  {Li}, {Challis}, {Kirshner}, {Matheson}, {Barris}, {Tonry}, {Riess},
  {Leibundgut}, {Sollerman}, {Spyromilio}, {Clocchiatti}, \&
  {Pompea}}]{essence-02}
{Smith}, R.~C., {et~al.} 2002{\natexlab{b}},
  American Astronomical Society Meeting, 201, 0

\bibitem[{{Spergel} {et~al.}(2003){Spergel}, {Verde}, {Peiris}, {Komatsu},
  {Nolta}, {Bennett}, {Halpern}, {Hinshaw}, {Jarosik}, {Kogut}, {Limon},
  {Meyer}, {Page}, {Tucker}, {Weiland}, {Wollack}, \&
  {Wright}}]{wmap-results03}
{Spergel}, D.~N., {et~al.} 2003, \apjs, 148, 175

\bibitem[{{Stokes} {et~al.}(2000){Stokes}, {Evans}, {Viggh}, {Shelly}, \&
  {Pearce}}]{linear}
{Stokes}, G.~H., {et~al.} 2000, Icarus, 148, 21

\bibitem[{{Sumi} {et~al.}(2003){Sumi}, {Abe}, {Bond}, {Dodd}, {Hearnshaw},
  {Honda}, {Honma}, {Kan-ya}, {Kilmartin}, {Masuda}, {Matsubara}, {Muraki},
  {Nakamura}, {Nishi}, {Noda}, {Ohnishi}, {Petterson}, {Rattenbury}, {Reid},
  {Saito}, {Saito}, {Sato}, {Sekiguchi}, {Skuljan}, {Sullivan}, {Takeuti},
  {Tristram}, {Wilkinson}, {Yanagisawa}, \& {Yock}}]{moa-bulge03}
{Sumi}, T., {et~al.} 2003, \apj, 591, 204

\bibitem[{{Tonry} {et~al.}(2003){Tonry}, {Schmidt}, {Barris}, {Candia},
  {Challis}, {Clocchiatti}, {Coil}, {Filippenko}, {Garnavich}, {Hogan},
  {Holland}, {Jha}, {Kirshner}, {Krisciunas}, {Leibundgut}, {Li}, {Matheson},
  {Phillips}, {Riess}, {Schommer}, {Smith}, {Sollerman}, {Spyromilio},
  {Stubbs}, \& {Suntzeff}}]{tonry-highz}
{Tonry}, J.~L., {et~al.} 2003, \apj, 594, 1

\bibitem[{{Tyson}(2002)}]{tyson-lsst02}
{Tyson}, J.~A. 2002, in Survey and Other Telescope Technologies and
  Discoveries. Edited by Tyson, J. Anthony; Wolff, Sidney. Proceedings of the
  SPIE, Volume 4836, pp. 10-20 (2002)., 10--20

\bibitem[{{Van Den Bergh}(2003)}]{vdb-gc}
{Van Den Bergh}, S. 2003, \apj, 590, 797

\bibitem[{{Vanden Berk} {et~al.}(2002){Vanden Berk}, {Lee}, {Wilhite},
  {Beacom}, {Lamb}, {Annis}, {Abazajian}, {McKay}, {Kron}, {Kent}, {Hurley},
  {Kehoe}, {Wren}, {Henden}, {York}, {Schneider}, {Adelman}, {Brinkmann},
  {Brunner}, {Csabai}, {Harvanek}, {Hennessy}, {Ivezi{\' c}}, {Kleinman},
  {Kleinman}, {Krzesinski}, {Long}, {Neilsen}, {Newman}, {Snedden},
  {Stoughton}, {Tucker}, \& {Yanny}}]{sloan-ot1}
{Vanden Berk}, D.~E., {et~al.} 2002, \apj, 576, 673

\bibitem[{{Vestrand} {et~al.}(2002){Vestrand}, {Borozdin}, {Brumby},
  {Casperson}, {Fenimore}, {Galassi}, {McGowan}, {Perkins}, {Priedhorsky},
  {Starr}, {White}, {Wozniak}, \& {Wren}}]{raptor}
{Vestrand}, W.~T., {et~al.} 2002, in Advanced Global Communications Technologies for Astronomy II.
  Edited by Kibrick, Robert I. Proceedings of the SPIE, Volume 4845, pp.
  126-136 (2002)., 126--136

\bibitem[{{Vivas} {et~al.}(2001){Vivas}, {Zinn}, {Andrews}, {Bailyn}, {Baltay},
  {Coppi}, {Ellman}, {Girard}, {Rabinowitz}, {Schaefer}, {Shin}, {Snyder},
  {Sofia}, {van Altena}, {Abad}, {Bongiovanni}, {Brice{\~ n}o}, {Bruzual},
  {Della Prugna}, {Herrera}, {Magris}, {Mateu}, {Pacheco}, {S{\' a}nchez},
  {S{\' a}nchez}, {Schenner}, {Stock}, {Vicente}, {Vieira}, {Ferr{\'{\i}}n},
  {Hernandez}, {Gebhard}, {Honeycutt}, {Mufson}, {Musser}, \&
  {Rengstorf}}]{vivas-quest}
{Vivas}, A.~K., {et~al.} 2001, \apjl, 554, L33

\bibitem[{{Voges} {et~al.}(2000){Voges}, {Aschenbach}, {Boller}, {Brauninger},
  {Briel}, {Burkert}, {Dennerl}, {Englhauser}, {Gruber}, {Haberl}, {Hartner},
  {Hasinger}, {Pfeffermann}, {Pietsch}, {Predehl}, {Schmitt}, {Trumper}, \&
  {Zimmermann}}]{rosat-faintcat}
{Voges}, W., {et~al.} 2000, VizieR Online Data Catalog, 9029, 0

\bibitem[{{White} {et~al.}(1997){White}, {Becker}, {Helfand}, \&
  {Gregg}}]{first-97}
{White}, R.~L., {et~al.} 1997, \apj, 475, 479

\bibitem[{{Wittman} {et~al.}(2000){Wittman}, {Becker}, {Margoniner}, {Tyson},
  {Dell'Antonio}, {Loomba}, {Schommer}, {Smith}, {Aussel}, {Sanders}, {Stern},
  {Dey}, {Dawson}, \& {Spinrad}}]{2000IAUC.7551....1W}
{Wittman}, D.~M., {et~al.} 2000, \iaucirc, 7551, 1

\bibitem[{{Wittman} {et~al.}(2002){Wittman}, {Tyson}, {Dell'Antonio}, {Becker},
  {Margoniner}, {Cohen}, {Norman}, {Loomba}, {Squires}, {Wilson}, {Stubbs},
  {Hennawi}, {Spergel}, {Boeshaar}, {Clocchiatti}, {Hamuy}, {Bernstein},
  {Gonzalez}, {Guhathakurta}, {Hu}, {Seljak}, \& {Zaritsky}}]{dls-spie02}
{Wittman}, D.~M., {et~al.} 2002, in Survey
  and Other Telescope Technologies and Discoveries. Edited by Tyson, J.
  Anthony; Wolff, Sidney. Proceedings of the SPIE, Volume 4836, pp. 73-82
  (2002)., 73--82

\bibitem[{{Wozniak} {et~al.}(2001){Wozniak}, {Udalski}, {Szymanski}, {Kubiak},
  {Pietrzynski}, {Soszynski}, \& {Zebrun}}]{ogle-events01}
{Wozniak}, P.~R., {et~al.} 2001, Acta Astronomica, 51, 175

\bibitem[{{Wozniak} {et~al.}(2004){Wozniak}}]{nsvs}
{Wozniak}, P.~R., {et~al.} 2004, \aj in press, astro-ph/0401217

\end{thebibliography}



\clearpage


\begin{deluxetable}{llll|llll}
\tablewidth{0pt}
\tabletypesize{\scriptsize}
\tablecaption{Unusual Transient Event Summary \label{tab-short} }
\tablehead{
  \colhead {Event} &
  \colhead {DLS Subfield} &
  \colhead {RA (J2000)} &
  \colhead {DEC (J2000)} &
  \colhead {$B$} &
  \colhead {$V$} &
  \colhead {$R$} &
  \colhead {$z'$} 
}
\startdata
OT 20020112 & F4p13 & 10:49:32.8 & -04:31:45  & $>27.3$         & $>27.2$         & $>27.7$         & $>24.9$         \\
\nodata      &       &            &            & $>27.2$         & $>27.0$         & $>27.6$         & $>24.8$         \\
&  & &  &  &  &  &  \\
OT 20010326 & F5p31 & 14:01:42.2 & -11:35:15  & $>26.4$         & $24.48 \pm 0.10$ & $23.30 \pm 0.12$   & $21.16 \pm 0.11$ \\
\nodata      &       &            &            & $>26.2$         & $24.27 \pm 0.10$ & $23.13 \pm 0.12$   & $21.06 \pm 0.11$ \\
&  & &  &  &  &  &  \\
OT 20020115 & F4p23 & 10:48:56.1 & -05:00:41  & $21.25 \pm 0.07$ & $19.67 \pm 0.05$ & $<19.5$        & $16.28 \pm 0.09$ \\
&  & &  &  &  &  &  \\
OT 20030305 & F4p31 & 10:53:45.8 & -05:37:44  & $>27.2$         & $>27.1$         & $24.58 \pm 0.10$ & $21.38 \pm 0.13$ \\
\nodata      &       &            &            & $>27.0$         & $>27.0$         & $24.50 \pm 0.10$ & $21.34 \pm 0.13$ \\
\enddata
\tablecomments{Summary of selected optical transients (OTs) from the
DLS transient survey.  OT 20020112 varied on the timescale of months,
and we fail to detect a host or precursor object to the reported
limiting magnitudes.  The other three events varied on
thousand--second timescales, and were resolved in multiple survey
images.  Magnitudes are Vega--based and represent any detected
precursor or host in its quiescent state.  A second set of magnitudes
takes into account \cite{schlegel} corrections for Galactic reddening,
assuming a $R_V = 3.1$ extinction curve, except for OT 20020115 which
is known to be Galactic in origin.  Point source limits from
Equation~\ref{eqn-limit} are quoted for null detections.}
\end{deluxetable}


\begin{deluxetable}{lllll}
\tablewidth{0pt}
\tabletypesize{\scriptsize}
\tablecaption{Optical Transient Event Photometry \label{tab-phot} }
\tablehead{ \colhead {Event} & \colhead {MJD} & \colhead {Passband} &
\colhead {$\mathcal{M}$} & \colhead {$\mathcal{M}_C$} } \startdata
OT 20020112 & 52285.3 & V & $22.24 \pm 0.11$ & $22.11 \pm 0.11$  \\
            & 52286.2 & B & $22.24 \pm 0.06$ & $22.07 \pm 0.06$  \\
            & 52287.2 & R & $22.10 \pm 0.08$ & $22.00 \pm 0.08$  \\
            & 52287.4 & V & $22.17 \pm 0.15$ & $22.04 \pm 0.15$  \\
            & 52288.2 & R & $22.23 \pm 0.04$ & $22.13 \pm 0.04$  \\
            & 52288.2 & B & $22.32 \pm 0.08$ & $22.15 \pm 0.08$  \\
            & 52288.3 & V & $22.18 \pm 0.11$ & $22.05 \pm 0.11$  \\
            & 52290.2 & B & $22.42 \pm 0.07$ & $22.25 \pm 0.07$  \\
            & 52290.3 & V & $22.31 \pm 0.11$ & $22.18 \pm 0.11$  \\
            & 52319.2 & V & $24.07 \pm 0.11$ & $23.94 \pm 0.11$  \\
            & 52321.3 & B & $25.46 \pm 0.14$ & $25.29 \pm 0.14$  \\
\hline
OT 20010326 & 51994.2315949 & B & $>24.8$         & $>24.4$         \\
            & 51994.2397245 & B & $22.76 \pm 0.06$ & $22.50 \pm 0.06$ \\
            & 51994.2478576 & B & $23.70 \pm 0.07$ & $23.44 \pm 0.07$ \\
            & 51994.2646898 & R & $>24.7$         & $>24.5$         \\
\hline
OT 20020115 & 52289.2963074 & B & $>25.0$         & \nodata  \\
            & 52289.3044363 & B & $20.73 \pm 0.07$ & \nodata \\
            & 52289.3125651 & B & $23.83 \pm 0.07$ & \nodata \\
            & 52290.2594999 & B & $>24.8$         & \nodata  \\
\hline
OT 20030305 & 52703.1732501 & V & $>23.8$         & $>23.7$         \\
            & 52703.1813791 & V & $20.99 \pm 0.12$ & $20.89 \pm 0.12$ \\
            & 52703.1895094 & V & $22.22 \pm 0.09$ & $22.12 \pm 0.09$ \\
            & 52703.1995017 & B & $22.50 \pm 0.08$ & $22.37 \pm 0.08$ \\
            & 52703.2076305 & B & $21.57 \pm 0.06$ & $21.44 \pm 0.06$ \\
            & 52703.2157589 & B & $21.49 \pm 0.06$ & $21.36 \pm 0.06$ \\
            & 52703.2249089 & B & $21.70 \pm 0.06$ & $21.57 \pm 0.06$ \\
            & 52703.2330383 & B & $21.77 \pm 0.07$ & $21.64 \pm 0.07$ \\
            & 52705.3304513 & V & $>23.9$         & $>23.8$
\enddata
%
\tablecomments{Differential flux magnitudes $\mathcal{M}$ for each
  optical transient, in the Vega system.  Also listed are the Modified
  Julian Day (MJD) and filter for each observation.  OT 20020112
  varied on a month-long timescale, and the photometry listed
  represents the MJD--averaged brightness.  The constraining
  observations preceding and following the event are also listed for
  the short timescale transients.  Limits on $\mathcal{M}$ are
  determined by adding input PSFs to the difference images.  Error
  bars include systematic calibration errors added in quadrature with
  the experimental uncertainties in the measurement.  $\mathcal{M}_C$
  represents \cite{schlegel} corrected magnitudes, assuming a source
  outside the Galactic extinction layer, except for OT 20020115 which
  is known to be Galactic in origin.}
\end{deluxetable} 


\begin{deluxetable}{llclllcccc}
\tabletypesize{\scriptsize}
\tablewidth{0pt}
\tablecaption{Event rate \label{tab-rate} }
\tablehead{
  \colhead {Filter} &
  \colhead {Timescale} &
  \colhead {Exposure~$\exposure$ (deg$^2$--days)} &
  \colhead {$\mathcal{M}_{min}$} &
  \colhead {$\mathcal{M}_{max}$} &
  \colhead {$\left< \eff \right>$} &
  \colhead {$N_{obs}$} &
  \colhead {$N_{max, 95\%}$} &
  \colhead {$\rate$} & 
  \colhead {$\rate_{max, 95\%}$} 
}
\startdata
B      & 1300s  & 1.1    & 18.6   & 23.8   &  0.65  & 3      & 7.8    & 6.5 & 17 \\
\nodata & \nodata & \nodata & \nodata & \nodata & \nodata & 2      & 6.3    & 4.3 & 14 \\
\nodata & \nodata & \nodata & \nodata & \nodata & \nodata & 1      & 4.7    & 2.2 & 10 \\
\nodata & \nodata & \nodata & \nodata & \nodata & \nodata & 0      & 3.0    & 0.0 & 6.5 \\
\hline
V      & 1300s  & 1.2    & 18.8   & 23.3   &  0.63  & 1      & 4.7    & 2.1 & 9.9 \\
\nodata & \nodata & \nodata & \nodata & \nodata & \nodata & 0      & 3.0    & 0.0 & 6.3 \\
\hline
R      & 1900s  & 1.5    & 19.5   & 23.4   &  0.62  & 0      & 3.0    & 0.0 & 5.2 \\
\hline
B+V+R  & \nodata & 3.7    & \nodata & \nodata &  0.63  & 4      & 9.2    & 2.7 & 6.3 \\
\nodata & \nodata & \nodata & \nodata & \nodata & \nodata & 3      & 7.8    & 2.0 & 5.3 \\
\nodata & \nodata & \nodata & \nodata & \nodata & \nodata & 2      & 6.3    & 1.4 & 4.3 \\
\nodata & \nodata & \nodata & \nodata & \nodata & \nodata & 1      & 4.7    & 0.7 & 3.2 \\
\nodata & \nodata & \nodata & \nodata & \nodata & \nodata & 0      & 3.0    & 0.0 & 2.0
\enddata
%
\tablecomments{For each filter, we present our sensitivity to
transients for timescales at twice the typical exposure times plus
$100s$ of camera readout.  We list the magnitude ranges over which our
efficiency of recovering a transient is approximated by a constant
$\left<\eff\right>$.  Assuming we have detected $0 \leq N \leq
N_{obs}$ events per filter, we also list the maximum number of events
allowed by Poisson statistics at the $95\%$ confidence level, for any
model of temporal variability.  We derive the corresponding event rate
$\rate$ in events deg$^{-2}~{\rm day}^{-1}$.}
\end{deluxetable} 


\end{document}